\PassOptionsToPackage{table,xcdraw,dvipsnames}{xcolor}
\PassOptionsToPackage{hyphens}{url}
\PassOptionsToPackage{unicode}{hyperref}
\PassOptionsToPackage{naturalnames}{hyperref}

 \documentclass[sigplan, screen, authorversion]{acmart}

\usepackage{enumitem}
\usepackage{multirow} 
\usepackage{adjustbox} 
\usepackage{subcaption} 
\usepackage{tabularx} 
\usepackage{tikz} 
\usepackage{fixltx2e} 

\usepackage{pifont}
\newcommand{\cmark}{\ding{51}}%
\newcommand{\xmark}{\ding{55}}%

\usetikzlibrary{shapes,arrows}
\usetikzlibrary{decorations.pathreplacing}

\newcommand{\beginbsec}[1]{\vspace{3pt}\noindent\textbf{\textit{#1 \hspace{2pt}}}}
\newcommand{\beginbseceval}[1]{\vspace{2pt}\noindent\underline{\textit{$\triangleright$ #1:}}}

\newcommand{\floor}[1]{\lfloor #1 \rfloor}

\newcommand{\CAP}[1]{\scalebox{0.85}{#1}}

\newcommand{\VAR}[1]{\texttt{#1}}

\def\cid{\texorpdfstring{\VAR{c\textsubscript{id}}}{}} 

\newcommand*\circled[1]{\tikz[baseline=(char.base)]{
            \node[shape=circle,draw,inner sep=0.5pt] (char) {#1};}}

\newcommand{\squishlist}{
 \begin{list}{$\bullet$}
  { \setlength{\itemsep}{2pt}
     \setlength{\parsep}{0pt}
     \setlength{\topsep}{2pt}
     \setlength{\partopsep}{0pt}
     \setlength{\leftmargin}{1em}
     \setlength{\labelwidth}{1em}
     \setlength{\labelsep}{0.5em} } 
}

\newcommand{\squishlistContrib}{ %
 \begin{list}{$\bullet$}
  { \setlength{\itemsep}{2pt}
     \setlength{\parsep}{0pt}
     \setlength{\topsep}{2pt}
     \setlength{\partopsep}{0pt}
     \setlength{\leftmargin}{1em}
     \setlength{\labelwidth}{1em}
     \setlength{\labelsep}{0.5em} }
}
\newcommand{\squishend}{ \end{list}  }

\def\eg{e.g.,~} 
\def\ie{i.e.,~} 

\copyrightyear{2020} 
\acmYear{2020} 
\setcopyright{acmcopyright}
\acmConference[ASPLOS '20]{Proceedings of the Twenty-Fifth International Conference on Architectural Support for Programming Languages and Operating Systems}{March 16--20, 2020}{Lausanne, Switzerland}
\acmBooktitle{Proceedings of the Twenty-Fifth International Conference on Architectural Support for Programming Languages and Operating Systems (ASPLOS '20), March 16--20, 2020, Lausanne, Switzerland} 
\acmPrice{15.00}
\acmDOI{10.1145/3373376.3378496}
\acmISBN{978-1-4503-7102-5/20/03}

\title[Hermes Replication Protocol]{Hermes: a Fast, Fault-Tolerant and Linearizable Replication Protocol}

\begin{document}

\author{Antonios Katsarakis, Vasilis Gavrielatos, M. R. Siavash Katebzadeh,}
\author{Arpit Joshi\texorpdfstring{$^{*}$}{*}, Aleksandar Dragojevic\texorpdfstring{$^{**}$}{**}, Boris Grot, Vijay Nagarajan}

\affiliation{
\institution{University of Edinburgh, $^{*}$Intel, $^{**}$Microsoft Research}
}

\renewcommand{\shortauthors}{A. Katsarakis, et al.}

%


%

\begin{abstract}
Today's datacenter applications are underpinned by datastores that are responsible for providing availability, consistency, and performance. For high availability in the presence of failures, these datastores replicate data across several nodes. This is accomplished with the help of a \emph{reliable replication protocol} that is responsible for maintaining the replicas strongly-consistent even when faults occur. Strong consistency is preferred to weaker consistency models that cannot guarantee an intuitive behavior for the clients. Furthermore, to accommodate high demand at real-time latencies, datastores must deliver high throughput and low latency. 

This work introduces Hermes\footnote{The name is inspired by the immortal Olympian figure, who was the messenger of the gods and a conductor of souls into the afterlife.}, a broadcast-based reliable replication protocol for in-memory datastores that provides both high throughput and low latency by enabling local reads and fully-concurrent fast writes at all replicas. Hermes couples \emph{logical timestamps} with cache-coherence-inspired \emph{invalidations} to guarantee linearizability,  avoid write serialization at a centralized ordering point, resolve write conflicts locally at each replica (hence ensuring that writes never abort) and provide fault-tolerance via replayable writes. Our implementation of Hermes over an \CAP{RDMA}-enabled reliable datastore with five replicas shows that Hermes consistently achieves higher throughput than state-of-the-art \CAP{RDMA}-based reliable protocols (\CAP{ZAB} and \CAP{CRAQ}) across all write ratios while also significantly reducing tail latency. At 5\% writes, the tail latency of Hermes is 3.6$\times$ lower than that of \CAP{CRAQ} and \CAP{ZAB}.
\end{abstract}

\maketitle


\vspace{-5pt}
\section{Introduction}
\label{sec:introduction}
Today's online services and cloud applications rely on high-performance datastores\footnote{We use the term datastore broadly to encompass a wide range of in-memory storage systems with an \CAP{API} for reading and writing objects (keys).}, such as key-value stores (\CAP{KVS}) and lock services, for storing and accessing their dataset.
These datastores must provide high throughput at very low latencies while offering high availability, as they are deployed on failure-prone commodity infrastructure~\cite{Forbes:2013}.
Keeping the dataset in-memory and exploiting high-performance datacenter networking (e.g., \CAP{RDMA}) is essential, but not sufficient.

Data replication is a fundamental feature of high performance and reliable datastores. Data must be replicated across multiple nodes to increase throughput 
because 
a single node often cannot keep up with the request load~\cite{Bronson:2013}. Replication is also necessary to guarantee that a failure of a node or a network link does not render a portion of the dataset inaccessible. 
Maintaining
the replicas strongly-consistent, to ensure that the services running on the datastore operate correctly, is a challenge, especially in the presence of failures. A \emph{
reliable
replication protocol} is responsible for 
keeping
the replicas of a datastore strongly-consistent~-- even when faults occur~-- by determining the necessary
actions for the execution of reads and writes. 

When it comes to performance, recent works on 
reliably-replicated datastores focus on 
throughput~\cite{Terrace:2009} and tend to ignore latency. Meanwhile, latency is emerging as a critical design goal in the age of interactive services and machine actors~\cite{Barroso:2017}. For instance, Anwar et al.~\cite{Anwar:2018} note that a deep learning system running on top of a reliable datastore is profoundly affected by the latency of the datastore.

Today's replication protocols are not designed to handle the latency challenge of in-memory reliable datastores. Chain Replication (\CAP{CR})~\cite{VanRenesse:2004}, a state-of-the-art high performance reliable replication protocol~\cite{Anwar:2018} is a striking example of trading latency for throughput. Our detailed study of \CAP{CRAQ}~\cite{Terrace:2009}, the state-of-the-art \CAP{CR} variant, reveals that whilst \CAP{CRAQ} can offer very high throughput, it is ill-suited for latency-sensitive workloads. \CAP{CRAQ} organizes the replicas in a chain. While reads can be served locally by each of the replicas, writes expose the entire length of the chain. Moreover, when a read hits a key for which a write is in progress, the read incurs an additional latency as it waits for the write to be resolved. With high-latency writes, and mixed-latency reads, \CAP{CRAQ} fails to provide predictably low latency.

This work addresses the challenge of designing a reliable replication protocol that provides both high throughput and low latency within a datacenter. To that end, we identify key features necessary for high performance, which are summarized in Table~\ref{tab:hp-features}.
For reads, this means the ability to
execute a read locally 
on any of the replicas. For writes, high performance mandates the ability to execute writes 
in a decentralized manner (i.e., any replica can initiate and drive a write to completion without serializing it through another node), 
concurrently execute writes to different keys, and  
complete writes fast
(e.g., by minimizing round-trips).

Based on these insights, we introduce Hermes, a strongly-consistent fault-tolerant replication protocol for in-memory datastores that provides high throughput and low latency. At a high level, Hermes is a broadcast-based protocol for single-key reads, writes and \CAP{RMW}s that resembles two-phase commit (\CAP{2PC})~\cite{Gray:1978}. However, \CAP{2PC} is not reliable (\S\ref{sec:related-work}) and is overkill for replicating single-key writes. In contrast, Hermes is highly optimized for single-key operations and is reliable.

Hermes combines two ideas to achieve high performance. The first is the use of {\em invalidations}, which is a form of light-weight locking inspired by cache coherence protocols. The second is per-key \textit{logical timestamps} implemented as Lamport clocks~\cite{Lamport:1978}. Together, these enable linearizability, local reads and fully-concurrent, decentralized, and fast writes. Logical timestamps further allow each node to locally establish a single global order of writes to a key, which enables conflict-free write resolution (i.e., writes never abort\footnote{Read-Modify-Writes (\CAP{RMW}s) in Hermes may abort (\S\ref{sec:hermes-RMWs}).} -- another difference from \CAP{2PC}) and 
\textit{write replays} to handle faults.


\begin{table}[t]
\centering

\begin{adjustbox}{max width=0.45\textwidth}

\begin{tikzpicture}
\renewcommand{\arraystretch}{1.5}
\node (table) [inner sep=-0pt] {
\begin{tabular}{
>{\columncolor[HTML]{C0C0C0}}c l}
\cellcolor[HTML]{C0C0C0}                                 & local         \\ \cline{2-2} 
\multirow{-2}{*}{\cellcolor[HTML]{C0C0C0}\textbf{reads}} & load-balanced
\end{tabular}
};
\draw [rounded corners=.3em] (table.north west) rectangle (table.south east);
\end{tikzpicture}

\begin{tikzpicture}
\node (table) [inner sep=0pt] {
\begin{tabular}{
>{\columncolor[HTML]{C0C0C0}}l l}
\cellcolor[HTML]{C0C0C0}                                  & decentralized                         \\ \cline{2-2} 
\cellcolor[HTML]{C0C0C0}                                  & inter-key concurrent                  \\ \cline{2-2} 
\multirow{-3}{*}{\cellcolor[HTML]{C0C0C0}\textbf{writes}} & fast (\eg few \CAP{RTT}s)
\end{tabular}
};
\draw [rounded corners=.3em] (table.north west) rectangle (table.south east);
\end{tikzpicture}
\end{adjustbox}

\caption{Replication protocol features for high-performance}
\label{tab:hp-features}
\vspace{-20pt}
\end{table}

\noindent To summarize, the contributions of this work are as follows:

\begin{itemize}[leftmargin=*]
    \item \scalebox{0.96}{\textbf{Introduces \textit{Hermes}, a reliable replication protocol} that} utilizes invalidations and logical timestamps to achieve high performance and linearizability. 
    Any replica in Hermes allows for efficient local reads and fast fully-concurrent writes.
Hermes handles message loss and node failures by guaranteeing that any write can always be safely replayed. 
    
    \item \textbf{Formally verifies \textit{Hermes}} 
    in \CAP{$TLA^{+}$}~\cite{Lamport:1994} for safety and absence of deadlocks in the presence of crash-stop failures, message reorderings and duplicates.
    
    \item \scalebox{0.96}{\textbf{Implements a high-performance \CAP{RDMA}-based reliable}}
    \textbf{\CAP{KVS}}
    incorporating Hermes with \textit{Wings},
    our efficient \CAP{RDMA} \CAP{RPC} library.
    Our evaluation of Hermes shows that it outperforms the state-of-the-art \CAP{RDMA}-enabled virtual Paxos~\cite{Jha:2019} 
    protocol by an order of magnitude.
    Moreover, Hermes achieves higher throughput than the highly-optimized \CAP{RDMA}-based state-of-the-art \CAP{ZAB}~\cite{Junqueira:2011} and \CAP{CRAQ}~\cite{Terrace:2009} replication protocols across all write ratios while significantly reducing the tail latency. At $5\%$ writes, the tail latency of Hermes is at least 3.6$\times$ lower than that of \CAP{CRAQ} and \CAP{ZAB}.
\end{itemize}

\section{Background}
\label{sec:background}
\subsection{In-Memory Distributed Datastores}

This work focuses on a replication protocol that can be deployed over datastores, replicated within a local area network such as a datacenter. Clients typically interact with a datastore by first establishing a session through which they issue read and write requests. These datastores keep the application dataset in-memory and employ efficient communication primitives (e.g., \CAP{RDMA} or \CAP{DPDK}) to achieve high throughput at very low latencies. 
One example of such datastores is key-value stores (\CAP{KVS})~\cite{DeCandia:2007, Lakshman:2010, Bronson:2013, Dragojevic:2014, Lim:2014} that serve as the backbone for many of today's data-intensive online services, including e-commerce and social networks. Another example is lock services, such as Apache Zookeeper~\cite{Hunt:2010} and Google's Chubby~\cite{Burrows:2006}, which provide an \CAP{API} to the clients that allows them to maintain critical data, including locks.

\subsection{Replication and Consistency}
\label{sec:background-replication}

Datastores typically partition the stored data into smaller pieces called \emph{shards} and replicate each shard to guarantee fault tolerance. A fault-tolerant replication protocol is then deployed to enforce consistency and fault tolerance across all replicas of a given shard. The number of replicas for a shard is the \emph{replication degree}, and it presents a trade-off between cost and fault tolerance: more replicas increase fault tolerance, but also increase the cost of the deployment. A replication degree between 3 to 7 replicas is commonly considered to offer a good balance between safety and cost~\cite{Hunt:2010}. Thus, although a datastore may span numerous nodes, the replication protocol need only scale with the replication degree.

Whenever data are replicated, a consistency model must be enforced. While weak consistency can be leveraged to increase performance, it can also lead to nasty surprises when developers or clients attempt to reason about the system's behavior~\cite{Vogels:2009}. For this reason, this work focuses on \textit{reliable replication protocols} that offer the strongest consistency model: \emph{Linearizability} (Lin)~\cite{Herlihy:1990}, which mandates that each request appears to take effect globally and instantaneously at some point between its invocation and completion.
Lin has intuitive behavior, is compositional, and allows for the broadest spectrum of applications~\cite{Herlihy:2008, Viotti:2016}. 

\subsection{High Performance}
\label{sec:high-perf}
Maintaining high performance under strong consistency and fault tolerance is an established challenge~\cite{VanRenesse:2004, Baker:2011}. In the context of in-memory datastores, high performance is accepted to mean low latency and high throughput. Requirements for achieving high performance differ for reads and writes.

\beginbsec{Reads}
The key to achieving both low latency and high throughput on reads is (1) being able to service a read on any replica, which we call {\em load-balanced reads}, and (2) completing the read locally (i.e., without engaging other replicas). 
%
While seemingly trivial, load-balanced local reads (referred to as just {\em local reads} from now on) are a challenge for many reliable protocols, which may require communication among nodes to agree on a read value (e.g., \CAP{ABD}~\cite{Attiya:1995,Lynch:1997} and Paxos~\cite{Lamport:1998}) or that mandate that only a single replica serve linearizable reads for a given key (e.g., Primary-backup~\cite{Alsberg:1976}).

\beginbsec{Writes}
Achieving high write performance under strong consistency and fault tolerance is notoriously difficult. We identify the following requirements necessary for low-latency high-throughput writes:

\noindent $\succ$ \underline{{\em Decentralized}}: In order to reduce network hops and preserve load balance across the replica ensemble, any replica must be able to initiate a write and drive it to completion (by communicating with the rest of the replicas) whilst avoiding centralized serialization points. For instance, both \CAP{ZAB} and \CAP{CR} require all writes to initiate at a particular node, hence failing to achieve decentralized writes.

\noindent $\succ$ \underline{{\em Inter-key concurrent}}: Independent writes on different keys should be able to proceed in parallel,
to enable intra- and multi-threaded parallel request execution.
For example, \CAP{ZAB} requires all writes to be serialized through a leader, thus failing to provide inter-key concurrency.

\noindent $\succ$ \underline{{\em Fast}}: Fast writes require minimizing the number of message round-trips, avoiding long message chains (e.g., in contrast to \CAP{CR}), and shunning techniques that otherwise increase write latency (e.g., 
performing writes in lock-step~\cite{Mao:2008,Poke:2017}).

\subsection{Reliable Replication Protocols}
\label{sec:reliable-protocols}
%

\beginbsec{Failure model}
We consider a partially synchronous system~\cite{Dwork:1988} where 
processes are equipped with loosely synchronized clocks
(\CAP{LSC}s)\footnote{
Some reliable replication protocols can maintain safety and liveness without \CAP{LSC}s.
We discuss one such variant of our Hermes protocol in \S\ref{sec:discussion}.}
and crash-stop or network failures may occur (as in~\cite{Chandra:2016}).
In this model, processes may fail by crashing and their operation is non-Byzantine. Additionally, network failures can manifest as either (1) message reordering, duplication and loss, or (2) link failures that may lead to network partitions.

Reliable replication protocols capable of dealing with failures under the above failure model can be classified into two categories: \textit{majority-based} protocols, which are typically variants of Paxos~\cite{Lamport:1998}, and protocols that require
a stable membership of live nodes (\textit{membership-based} protocols).

\beginbsec{Majority-based protocols}
This class of protocols requires the majority of nodes to respond in order to commit a write, making it naturally tolerant to failures provided that a majority is responsive.
However, majority-based protocols pay the price in performance since -- in the absence of responses from all replicas -- there is no guarantee that a given write has reached all replicas, which makes linearizable local reads fundamentally challenging. Thus, 
most majority-based protocols give up on local reads but may support decentralized or inter-key concurrent writes~\cite{Lamport:1998, Lynch:1997, Moraru:2013}. Majority-based protocols that allow for local reads either
serialize independent writes on a master (e.g., \CAP{ZAB}) or 
require communication-intensive per-key leases (\S\ref{sec:related-work}); 
problematically, both ap\-proach\-es hurt 
performance
even in the absence of faults.

\begin{figure}[t]
  \centering
  \includegraphics[width=0.475\textwidth]{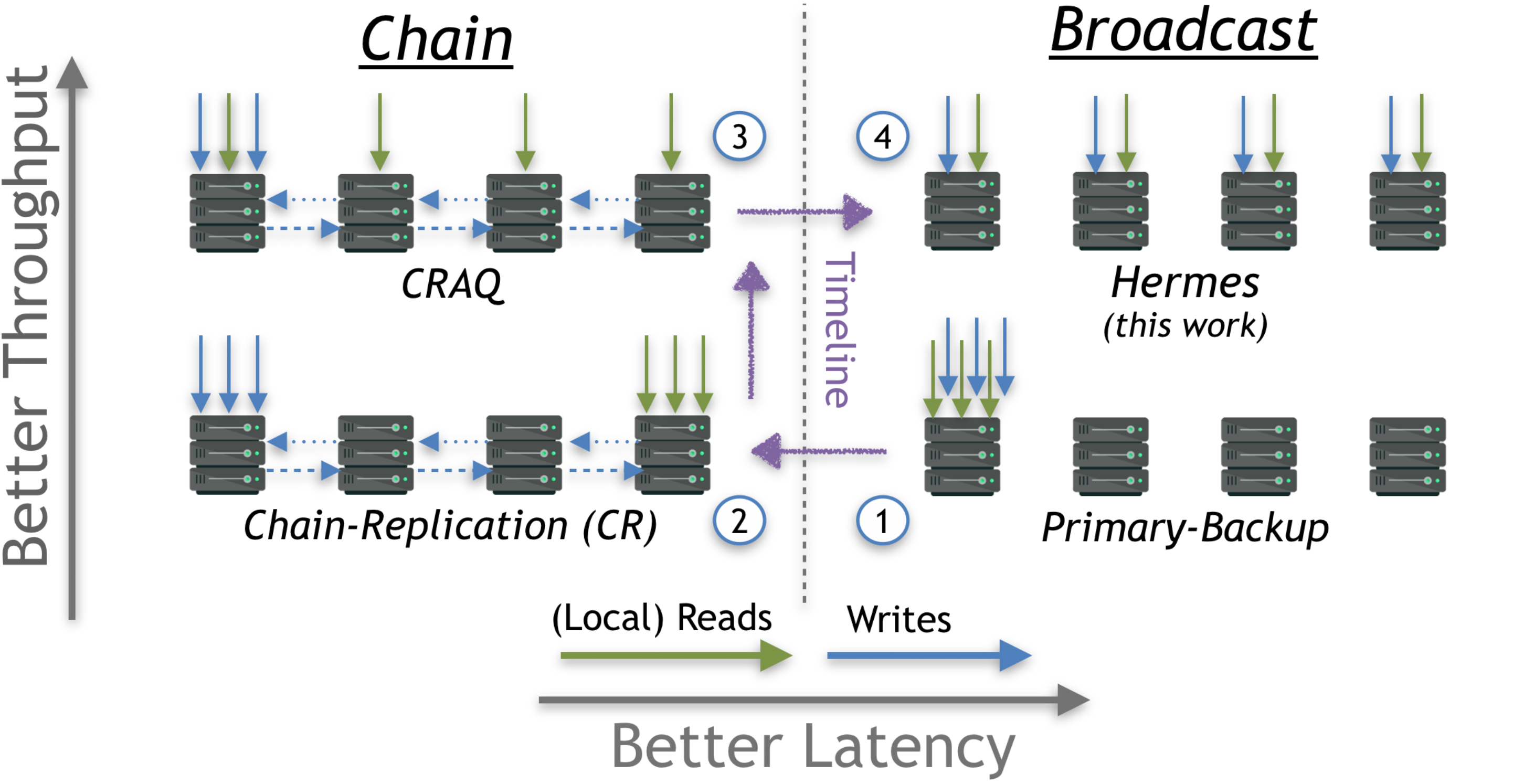}
  \vspace{-2.2em}
	\caption{Comparison of reliable membership-based protocols in terms of throughput and latency.} 
  \vspace{-15pt}
  \label{fig:design}
\end{figure}

\beginbsec{Membership-based protocols}
Protocols in this class require {\em all operational} 
nodes in the replica group to acknowledge each write (i.e., read-one/write-all protocols~\cite{Jimenez:2003}). In doing so, they assure that a committed write has reached all replicas in the ensemble, which naturally facilitates local reads without necessarily hindering write performance. 
Thus, in the absence of faults, membership-based protocols are naturally free of performance limitations associated with majority-based protocols.

Membership-based protocols are supported by a \textit{reliable membership} (\CAP{RM})~\cite{Kakivaya:2018}, typically based on Vertical Paxos~\cite{Lamport:2009}. Vertical Paxos uses a majority-based protocol to reliably maintain a stable membership of \textit{live} nodes~\cite{VanRenesse:1995} (i.e., as in virtual synchrony~\cite{Birman:1987}), which is guarded by leases.
Informally, nodes in Vertical Paxos
locally store a lease, a membership variable and an $epoch\_id$.
Nodes are \textit{operational} as long as their lease is valid.
Messages are tagged with the epoch\_id of the sender at the time of message creation, and a receiver drops any message tagged with a different epoch\_id than its local epoch\_id.
The membership variable establishes the set of live nodes, which allows for efficient execution of reads and writes on any node with a valid lease. During failure-free operation, membership leases are regularly renewed. 
When a failure is suspected, the membership variable is updated reliably (and epoch\_id is incremented) through a majority-based protocol but only after the expiration of leases. 
This circumvents potential 
false-positives of unreliable failure detection and 
maintains safety under network partitions (\S\ref{sec:hermes-discussion}). Simply put, updating the membership variable only after lease expiration ensures that unresponsive nodes 
have stopped serving requests before they are removed from the membership 
and new requests complete only amongst the remaining live nodes of the updated membership group.




A common practice for high-performance replication 
is to optimize for the typical failure-free operation by harnessing the performance benefits of membership-based protocols  
and limiting the usage of majority-based protocols to \CAP{RM} reconfiguration~\cite{Lamport:2009, Jimenez:2003, Dragojevic:2015}.
In fact, major datacenter operators, such as Microsoft, not only exploit membership-based protocols in their datastores~\cite{Dragojevic:2015, Shamis:2019}, but they also provide \CAP{LSC}s~\cite{Microsoft-time:2018, Corbett:2013} 
and \CAP{RM}~\cite{Kakivaya:2018} as datacenter services to ease the deployment of membership-based protocols by third parties. 

One of the earliest membership-based protocols is Primary-backup \cite{Alsberg:1976}, which serves all requests at a primary node and does not leverage the backup replicas for performance. 
Chain Replication \CAP{(CR)}~\cite{VanRenesse:2004} improves upon Primary-backup by organizing the nodes in a chain and dividing the responsibilities of the primary amongst the \textit{head} and the \textit{tail} of the chain, as shown in Figure~\ref{fig:design} (bottom-left). 
\CAP{CR} is a common choice for implementing high performance reliable replication~\cite{Jin:2018, Anwar:2018, Terrace:2009, Balakrishnan:2012, Wei:2017}.
We next discuss \CAP{CRAQ}~\cite{Terrace:2009}, a highly optimized variant of \CAP{CR}.

\subsection{CRAQ} 
\label{sec:craq} 

\CAP{CRAQ} is a state-of-the-art membership-based protocol that offers high throughput and strong consistency (Lin). In \CAP{CRAQ}, nodes are organized in a chain and writes are directed to its head, as in \CAP{CR}. The head propagates the write down the chain, 
which completes 
once it reaches the tail. Subsequently, the tail propagates acknowledgment messages upstream towards the head, letting all nodes know about the write's completion.

\CAP{CRAQ} improves upon \CAP{CR} by enabling read requests to be served locally from all nodes, as shown in Figure~\ref{fig:design} (top-left). However, if a non-tail node is attempting to serve a read for which it has seen a write message propagating downstream from head to tail, but has not seen the acknowledgement propagating up, then the tail must be queried to find out whether the write has been applied or not. 

\CAP{CRAQ} is the state-of-the-art reliable replication protocol that achieves high throughput via a combination of local reads and inter-key concurrent writes. However, \CAP{CRAQ} fails to satisfy the low latency requirement: while reads are typically local and thus very fast,  writes must traverse multiple nodes sequentially incurring a prohibitive latency overhead.

\section{Hermes} 
\label{sec:hermes}
Hermes
is a reliable mem\-ber\-ship-based broadcasting protocol that offers high throughput and low latency
whilst providing linearizable reads, writes, and \CAP{RMW}s (single-key transactions). 
Hermes optimizes for the common case of no failures~\cite{Barroso:2018} and targets intra-datacenter in-memory datastores 
with a replication degree typical of today's deployments (3-7 replicas)~\cite{Hunt:2010}.
As noted in \S\ref{sec:background-replication}, the replica count does not constrain the size of 
a sharded datastore, since each
shard is
replicated independently of other shards.
%
Example applications that can benefit from Hermes include reliable datastores~\cite{Baker:2011, Balakrishnan:2012, Edmund:2012, Wei:2017}, lock-services~\cite{Hunt:2010,Burrows:2006} and applications that require high performance, strong consistency and availability (e.g.,~\cite{Adya:2016,Botelho:2013,Woo:2018}).

\definecolor{myOrange}{RGB}{204, 102, 0}
\definecolor{myBlue}{RGB}{0, 114, 178}

\tikzstyle{sline} = [draw, -] 
\tikzstyle{arrow} = [draw, -latex'] 
\tikzstyle{client} = [rectangle, draw, fill=gray!30, 
    minimum width=1em, text centered,  minimum height=1em, node distance = 2em]
\tikzstyle{server} = [rectangle, draw, fill=blue!20, 
    minimum width=1em, text centered,  minimum height=0.7em, node distance = 1.5em]
\tikzstyle{dashed_rectangle} = [rectangle, draw, dashed, 
    text width=3.8em, text centered, rounded corners, minimum height=1.4em]
\tikzstyle{session} = [draw, text width = 2.5em, rounded corners,node distance = 1em, xshift = 4em]

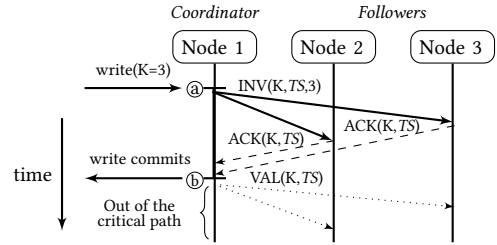
\begin{figure}[t]
\centering
\begin{tikzpicture}[auto]
    \path[arrow, thick] (-0.8,-1.3) -- node[left] {\footnotesize time} ++ (0,-1.5);
    \node[session, xshift = -0.5em, yshift = -1em] (node1) {\scalebox{0.8}{Node 1}};
    
    \node[session, xshift = -0.5em, right of = node1] (node2) {\scalebox{0.8}{Node 2}};
    \node[session, xshift = -0.5em, right of = node2] (node3) {\scalebox{0.8}{Node 3}};
    
    \node[left of = node1, yshift=-6em] (o1) {\scalebox{0.65}{Out of the}};
    \node[below of = o1, yshift=2.2em] (o2) {\scalebox{0.65}{critical path}};
    
    \draw [right of = o1, decorate, decoration={brace,amplitude=3pt},xshift=-1em,yshift=-9.7em]
    (0.5,0.5) -- (0.5,1.2) node [black,midway,xshift=-2em] {};

    \node[above of = node1, yshift=-4.45em, xshift = -0.75em] (ref1){\scalebox{0.75}{\circled{a}}};
    \node[above of = node1, yshift=-7.9em, xshift = -0.75em] (ref2){\scalebox{0.63}{\circled{b}}};
    
    \node[above of = node1, yshift=-1.5em] (coord){\scalebox{0.7}{\textit{Coordinator}}};
    \node[above of = node2, xshift = 2.2em, yshift=-1.5em] (followers){\scalebox{0.7}{\textit{Followers}}};
    
    \path[sline, thick]  (node1) -- node[left] {} ++ (0,-2.6);
    \path[sline, thick]  (node2) -- node[left] {} ++ (0,-2.6);
    \path[sline, thick]  (node3) -- node[left] {} ++ (0,-2.6);
    
    \path[arrow, thick] (-0.5, -0.9) -- node[above, xshift = 0em] {\scalebox{0.65}{write(K=3)}} ++ (1.3, 0);
    \path[arrow, thick] (0.8, -2.10) -- node[above, yshift = 0.15em, xshift = 0.25em] {\scalebox{0.65}{write commits}} ++ (-1.3, 0);
    \path[sline, thick] (1.38, -0.9) -- ++ (-0.3, 0);
    \path[sline, thick] (1.38, -2.10) -- ++ (-0.3, 0);
    \path[sline, thick] (1.215, -0.9) -- ++ (0, -1.2);
    
    \path[arrow, thick] (1.20, -0.95) -- node[above] {} ++ (1.6,-0.65);
    \path[arrow, thick] (1.20, -0.95) -- node[above, yshift = 0.1em, xshift = -2em] {\scalebox{0.65}{INV(K,\textit{TS},3)}} ++ (3.2,-0.4);
    
    \path[arrow, dashed] (2.82,-1.60) -- node[above, yshift = -0.2em, xshift = -0.3em] {\scalebox{0.65}{ACK(K,\textit{TS})}} ++ (-1.6, -0.3);
    \path[arrow, dashed] (4.42,-1.40) -- node[above, yshift=0.25em, xshift = 1.8em] {\scalebox{0.65}{ACK(K,\textit{TS})}} ++ (-3.2, -0.65);
    
    
    \path[arrow, dotted] (1.22, -2.18) -- node[below] {} ++ (1.6, -0.6);
    \path[arrow, dotted] (1.22, -2.18) -- node[above, xshift = -1.8em] {\scalebox{0.65}{VAL(K,\textit{TS})}} ++ (3.2, -0.3);
    
    
    
\end{tikzpicture}
\vspace{-1.em}
\caption{Example of writing a value of 3 to key K. Nodes one, two and three hold a replica of K. {\em {TS}} is the timestamp.}
\label{fig:Hermes-write}
\vspace{-15pt}
\end{figure}

\subsection{Overview}

In Hermes, reads complete locally. Writes can be initiated by any replica and complete fast regardless of conflicts. 
As illustrated in Figure~\ref{fig:Hermes-write}, a write to a key proceeds as follows: the replica initiating the write (called {\em coordinator}) broadcasts an {\em Invalidation (\CAP{INV})} message to the rest of the replicas (called {\em followers}) and waits on {\em acknowledgments (\CAP{ACK}s)}. Once all \CAP{ACK}s have been received; the write completes via a {\em Validation (\CAP{VAL})} message broadcast by the coordinator replica. 

We now briefly overview the salient features of Hermes and discuss the specifics in the following subsections. 

\beginbsec{Invalidations} 
When an \CAP{INV} message is received, the target key is placed in an Invalid state, meaning that reads to the key cannot be served. While conceptually similar to a lock (e.g., in \CAP{2PC}), the key difference is that with invalidations, concurrent writes
to the same key do not fail and are resolved in place through the use of logical timestamps as discussed below. The use of invalidations is inspired by cache coherence protocols, where a cache line in an Invalid state 
informs the readers that they must wait for an updated value. 

\beginbsec{Logical timestamps} 
Each write in Hermes is tagged with a monotonically-increasing per-key logical timestamp, implemented using Lamport clocks~\cite{Lamport:1978} and computed locally at the coordinator replica. The timestamp is a lexicographically ordered tuple of 
[\VAR{v}, \cid]
combining a key's version number (\VAR{v}), which is 
incremented on every write, with the node id of the coordinator 
(\cid).
Two or more writes to a key are \textit{concurrent} if 
their execution is initiated by different replicas holding the same timestamp.
Non-concurrent writes to a key are ordered based on their timestamp version, while concurrent writes from different coordinators (same version) are ordered via their 
\texorpdfstring{\VAR{c\textsubscript{id}}}{}\footnote{
  More precisely, a timestamp A: [\texorpdfstring{\VAR{v\textsubscript{A}}}{}, \texorpdfstring{\VAR{c\textsubscript{idA}}}{}] is higher than a timestamp B: [\texorpdfstring{\VAR{v\textsubscript{B}}}{}, \texorpdfstring{\VAR{c\textsubscript{idB}}}{}], if either $\texorpdfstring{\VAR{v\textsubscript{A}}}{} > \texorpdfstring{\VAR{v\textsubscript{B}}}{}$ or $\texorpdfstring{\VAR{v\textsubscript{A}}}{} = \texorpdfstring{\VAR{v\textsubscript{B}}}{}$ and $\texorpdfstring{\VAR{c\textsubscript{idA}}}{} > \texorpdfstring{\VAR{c\textsubscript{idB}}}{}$.
}. 
Uniquely tagged writes allow each node to locally establish a global order of writes 
to a key.

\beginbsec{High-performance non-conflicting writes}
\label{sec:conc-writes}Hermes allows for high-performance writes (\S\ref{sec:high-perf}) by maximizing concurrency while maintaining low latency.
First, writes in Hermes are executed from any replica in a decentralized manner, eschewing the use of a serialization point (e.g., a leader); thus reducing the number of network hops and ensuring load balance. 
In contrast to approaches that globally order independent writes for strong consistency (e.g., \CAP{ZAB} -- \S\ref{sec:ZAB}), Hermes allows writes to different keys to proceed in parallel, hence achieving inter-key concurrency. 
This is accomplished via Hermes' 
approach
of invalidating all operational replicas 
to achieve linearizability.
When combined with the logical timestamps, invalidations permit concurrent writes to the same key to be correctly linearized at the endpoints; thus, writes do not appear to conflict, making aborts unnecessary. 

Finally, in the absence of a failure, writes in Hermes 
cost
one and a half round-trips (\CAP{INV$\rightarrow$ACK$\rightarrow$VAL}); however, the exposed latency is just a single round-trip for each node.
From the perspective of the coordinator, once all \CAP{ACK}s are received, it is safe to respond to a client because 
at this point, the write is guaranteed to be visible to all live replicas, and any future read cannot return the old value (\ie the write is \textit{committed} -- Figure~\ref{fig:Hermes-write}\scalebox{0.7}{\circled{b}}).
The followers also observe only a single round-trip (further optimized in \S\ref{sec:optimizations}), which starts once an \CAP{INV} arrives; at that point, each follower responds with an \CAP{ACK} and completes the write when a \CAP{VAL} is received.



\beginbsec{Safely replayable writes}
Node and network faults
during
a write to a key may leave the key in a permanently Invalid state in some or all of the nodes. To prevent this, Hermes allows any invalidated operational replica to replay the write to completion without violating linearizability. This is accomplished using two mechanisms. First, the new value for a key is propagated to the replicas in \CAP{INV} messages (Figure~\ref{fig:Hermes-write}\scalebox{0.85}{\circled{a}}).
Such \textit{early value propagation}
guarantees that every invalidated node is aware of the new value. Secondly, logical timestamps enable a precise global ordering of writes in each of the replicas. By combining these ideas, a node that finds a key in an Invalid state for an extended period 
can safely replay a write by taking on a coordinator role and retransmitting \CAP{INV} messages to the replica ensemble with the {\em original} timestamp (i.e., original version number and 
\cid),
hence preserving the global write order.

The above features afford the following properties:

\noindent $\succ$ \underline{{\em Strong consistency}}:
By invalidating all replicas of a key at the start of a write, Hermes ensures that a key in a Valid state is guaranteed to hold the most up-to-date value. Hermes enforces the invariant that a read may complete if and only if the key is in a Valid state, which provides linearizability. 

\noindent $\succ$ \underline{{\em High performance}}:
Local reads in concert with high performance broadcast-based non-conflicting writes from any replica help ensure both low latency and high throughput. 

\noindent $\succ$ \underline{{\em Fault tolerance}}:
Hermes uses safely replayable
writes to tolerate a range of faults, including message loss, node failures, and network partitions. As a membership-based protocol, Hermes is aided by \CAP{RM} to provide a stable group membership of live nodes in the face of failures and network partitions.

\subsection{Hermes Protocol in Detail}
\label{sec:protocol}
\begin{figure}[t]
  \centering
  \includegraphics[width=0.43\textwidth]{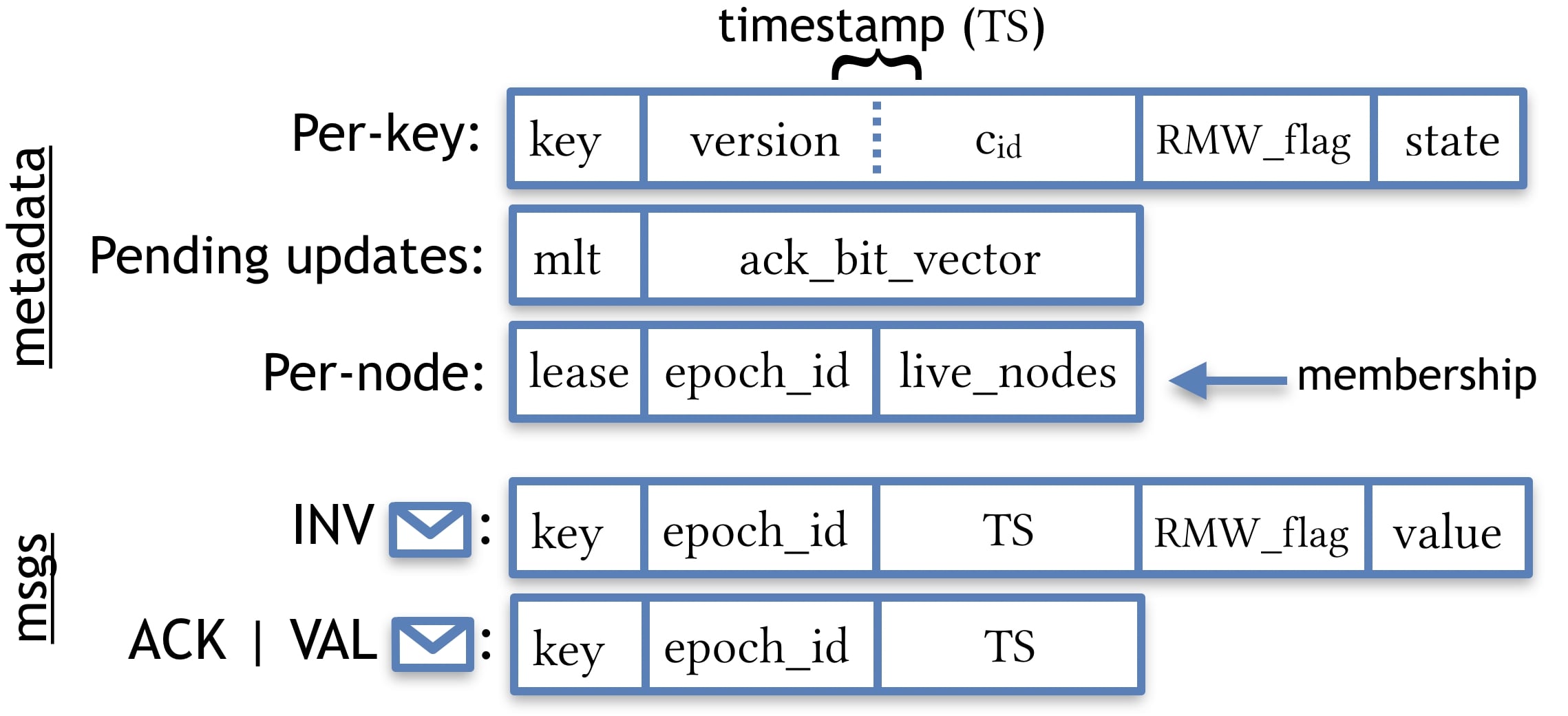}
  \vspace{-5pt}
  \caption{Metadata stored and messages sent by Hermes.}
  \vspace{-10pt}
  \label{fig:metadata}
\end{figure}

Hermes protocol consists of four stable states \textit{Valid}, \textit{Invalid}, \textit{Write} and \textit{Replay} and a single transient state \textit{Trans}.
%
Figure~\ref{fig:metadata} illustrates the format of protocol messages and the metadata stored at each replica.
A detailed protocol transition table, as well as the \CAP{$TLA^{+}$} specification, are available online\footnote{\href{https://hermes-protocol.com}{https://hermes-protocol.com}}.

The following protocol is slightly simplified in that it only focuses on reads and writes (omits \CAP{RMW}s) and only deals with node failures (but not network faults). 
Resilience to network faults and \CAP{RMW}s are described in \S\ref{sec:hermes-discussion} and \S\ref{sec:hermes-RMWs}, respectively.

\vspace{0.02in}
\noindent\textbf{\textit{Reads}}: A read request is serviced on an \textit{operational} replica (\ie one with an \CAP{RM} lease) by returning the local value of the requested key if it is in the Valid state. If the key is in any other state, the request is stalled.

\noindent\textbf{\textit{Writes}}:\\
\centerline{\underline{Coordinator}}
A coordinator node issues a write to a key only if it is in the Valid state; otherwise the write is stalled. To issue and complete a write, the coordinator node: 
\begin{itemize}[leftmargin=*]
\item\textbf{C\textsubscript{\CAP{TS}}}: Updates the key's local timestamp by incrementing its \VAR{version} and appending its node id as the 
\cid,
and assigns this new timestamp to the write.
\item\textbf{C\textsubscript{\CAP{INV}}}: Prompt\-ly broadcasts an \CAP{INV} message consisting of the \VAR{key}, the new timestamp (\VAR{\CAP{TS}}) and the \VAR{value} to all replicas and transitions the key to the Write state, whilst applying the new value locally. 
\item\textbf{C\textsubscript{\CAP{ACK}}}: Once the coordinator receives \CAP{ACK}s from all the \textit{live} replicas, the write is completed by transitioning the key to the Valid state (Invalid state if the key was in Trans state\footnote{The Trans state indicates a coordinator with a pending write that got invalidated. While not required, the Trans state is useful for tracking when the coordinator's original write completes, hence allowing the coordinator to notify the client of the write's completion.}).
\item\textbf{C\textsubscript{\CAP{VAL}}}: Finally, the coordinator broadcasts a \CAP{VAL}
consisting of the key and the same timestamp to all the followers.
\end{itemize}

\noindent Note that the coordinator waits for \CAP{ACK}s only from the live replicas 
as indicated in the membership variable. If a follower fails after an \CAP{INV} has been sent, the coordinator waits for the \CAP{ACK} from the failed node until the membership is reliably updated (after the node is detected as failed and the membership lease expires -- \S\ref{sec:reliable-protocols}). 
Once the coordinator is not missing any more \CAP{ACK}s, it can safely continue the write.

\vspace{0.02in}
\centerline{\underline{Follower}}
\begin{itemize}[leftmargin=*]
\item\textbf{F\textsubscript{\CAP{INV}}}: Upon receiving an \CAP{INV} message, a follower compares the timestamp from the incoming message to its local timestamp of the key. If the received timestamp is higher than the local timestamp, the follower transitions the key to the Invalid state (Trans state if the key was in the Write or the Replay state) and updates the key's local timestamp (both its \VAR{version} and \cid)
and value. 
\item \textbf{F\textsubscript{\CAP{ACK}}}: Irrespective of the result of the timestamp comparison, a follower always responds with an \CAP{ACK} 
containing
the same timestamp as that in the \CAP{INV} message of the write. 
\item \textbf{F\textsubscript{\CAP{VAL}}}: When a follower receives a \CAP{VAL} message, it transitions the key to the Valid state if and only if the received timestamp is equal to the key's local timestamp. Otherwise, the \CAP{VAL} message is simply ignored. 
\end{itemize}

\noindent\textbf{\textit{Write Replays}}:
A request that finds a key in the Invalid state for an extended period of time (determined via the {\em mlt} timer, described in~\S\ref{sec:hermes-discussion}) triggers a write replay. The node servicing the request takes on the coordinator role, transitions the key to the Replay state and begins a write replay by re-executing steps \textbf{C\textsubscript{\CAP{INV}}} through \textbf{C\textsubscript{\CAP{VAL}}} 
using the \CAP{TS} and value received with the \CAP{INV} message.
Note that the original \CAP{TS} is used in the replay (\ie the
\cid
is that of original coordinator) to allow the write to be correctly linearized.  
Once the replay is completed, the key transitions to the Valid state after which the initial request is serviced.

\noindent \textbf{$\rhd$} \textbf{Formal verification}: 
We expressed Hermes in \CAP{$TLA^{+}$}~\cite{Lamport:1994} and model checked the protocol's reads, writes, \CAP{RMW}s and replays for safety and absence of deadlocks in the presence of 
message reorderings and duplicates, and membership reconfigurations due to crash-stop failures.

\subsection{Hermes Protocol Optimizations}
\label{sec:optimizations}
\noindent\textbf{[O\textsubscript{1}] \textit{Eliminating unnecessary validations}} \hspace{2pt} When the coordinator of a write gathers all of its \CAP{ACK}s 
but discovers a concurrent write to the same key with a higher timestamp (i.e., was in the Trans state), it does not need to broadcast \CAP{VAL} messages (\textbf{C\textsubscript{\CAP{VAL}}}), thus saving valuable network bandwidth. 

\noindent\textbf{[O\textsubscript{2}] \textit{Enhancing fairness}} \hspace{2pt} Hermes linearizes writes based on their unique timestamps, consisting of a version and a node id.
In case of same versions (i.e., concurrent writes), the linearization is resolved based on the node ids, which might raise concerns about fairness. 
This is easily mitigated by assigning several \textit{virtual node ids} to each physical node. With this scheme, before issuing a write, a node randomly picks one of its assigned virtual node ids to be used for the write's logical timestamp. Of course, to maintain correctness, the same virtual node id cannot be assigned to more than one physical node. 
For example, given three nodes (\textit{A, B, and C}), the following sets of virtual ids \textit{A}:$\{1,4,7,10\}$, \textit{B}:$\{2,5,8,11\}$, and \textit{C}:$\{3,6,9,12\}$ are safe and would increase fairness. 

\noindent\textbf{[O\textsubscript{3}] \textit{Reducing blocking latency}} \hspace{2pt}
In the failure-free case, and during a write to a key,
followers block reads to that key for up to a round-trip (\S\ref{sec:conc-writes}).
This blocking latency can be reduced to a half round-trip if followers broadcast \CAP{ACK}s to all replicas instead of just responding to the coordinator of the write (\textbf{F\textsubscript{\CAP{ACK}}}). 
Once all \CAP{ACK}s have been received by a follower, it can service the reads to that key without waiting for the \CAP{VAL} message. While this optimization increases the number of \CAP{ACK}s, the actual bandwidth cost is minimal as \CAP{ACK} messages have a small constant size. The bandwidth cost is further offset by avoiding the need to broadcast \CAP{VAL} messages.
Thus, under the typical small replication degrees, this optimization comes at negligible cost in bandwidth.

\vspace{-2pt}
\subsection{Network Faults, Reconfiguration and Recovery}
\vspace{-1pt}
\label{sec:hermes-discussion}
This section presents Hermes' operation under imperfect links, network partitions
and the transient period of membership reconfiguration on a fault. 
It then 
provides an overview of the mechanism to add new nodes to the replica group.

\beginbsec{Imperfect Links}
In typical multi-path datacenter networks, messages can be reordered, duplicated, or lost~\cite{Farring:2009,Gill:2011,Lu:2018}. Hermes operates correctly under all of these scenarios as described below. In Hermes, the information necessary to linearize operations is embedded with the keys and in the messages in the form of logical timestamps. Thus, even if messages get delayed, reordered, or duplicated in the network, the protocol never violates linearizability.

Hermes uses the same idea of replaying writes if any of its \CAP{INV}, \CAP{ACK}, or \CAP{VAL} messages is suspected to be lost. A message is suspected to be lost for a key if the request's \textit{message-loss timeout (mlt)}, within which every write request is expected to be completed, is exceeded. To detect the loss of an \CAP{INV} or \CAP{ACK} for a particular write, the coordinator of the write resets the 
request's mlt 
once it broadcasts \CAP{INV} messages. If the
mlt
of a key is exceeded before its write completion, then the coordinator suspects a potential message loss and 
resets the request's mlt before retransmitting the write's \CAP{INV} broadcast.

In contrast, the loss of a \CAP{VAL} message is handled by the follower using a write replay. Once a follower receives a request for a key in the Invalid state, it resets the request's message-loss timeout.
If the timestamp or the state has not been updated within the 
mlt
duration, it suspects the loss of a \CAP{VAL} message and triggers a write replay. Although a write replay will never compromise the safety of the protocol, we note that a carefully calibrated timeout will reduce unnecessary replays (e.g., when messages are not lost).

\beginbsec{Network Partitions}
Datacenter network topologies are highly redundant~\cite{Gill:2011,Singh:2015}; however, in rare cases, link failures might result in a network partition. According to the \CAP{CAP} theorem~\cite{Brewer:2000,Gilbert:2002},
either consistency or availability must be sacrificed in the presence of network partitions. 
Hermes follows the guidelines of Brewer~\cite{Brewer:2012} to permit the datastore to continue serving requests only in its \textit{primary partition}, which is a partition with the majority of replicas. 
Although failure detectors cannot differentiate between node failures and network partitions,
the 
membership 
can
only
be reliably updated
in the primary partition -- due to its majority-based protocol -- and does so only after the expiration of the membership leases. Thus, replicas in a minority partition stop serving requests before the membership is updated and new requests are able to complete only in the primary partition.
While this approach allows the datastore to continue operating even under network partitions, it reduces Hermes resilience from $n-1$ node failures to tolerating less than $\floor{\frac{n}{2}}$ failures, if the \CAP{RM} protocol is run by the datastore replicas and not external nodes.
Nevertheless, this cost is similar to any other reliable protocol that tolerates network partitions~\cite{Jha:2019, Hunt:2010, Lamport:1998}. 
Once network connectivity is restored, nodes previously on a minority side can re-join the replica group via a recovery procedure explained below. 


\beginbsec{Membership reconfiguration after a failure} Following a network partition or a node failure and expiration of the leases for all of the nodes in a membership group, a majority-based protocol is used to reliably update the membership. We  refer to this update as \textit{m-update}, which consists of a lease renewal, a new list of live nodes and an incremented epoch\_id. Although the m-update is consistent even in the presence of faults, the update does not reach all live replicas instantaneously. 
Rather, there is a transient period when some replicas that are considered live, according to the latest value of the membership, have received the m-update while others have not and are still non-operational.

Hermes seamlessly deals with the transition of m-update without violating safety. Hermes' replicas which have received the m-update are able to act as coordinators and serve new requests. Thus, reads that find the target key in the Valid state can immediately be served as usual.
In contrast, writes or reads that require a replay (\ie targeted key is Invalid) are effectively stalled 
until all live nodes as indicated by the membership variable receive the m-update. This is because writes and write replays do not commit until all live replicas become operational and acknowledge their \CAP{INV} messages.

During this transition period, any live follower that has not yet received the latest m-update will simply drop the \CAP{INV} messages, because those messages are tagged with an \VAR{epoch\_id} greater than the follower's local \VAR{epoch\_id}. This manifests as a simple message loss to a coordinator which triggers retransmission of the \CAP{INV}s (\S\ref{sec:hermes-discussion}). The coordinator eventually completes its writes once all live followers have received the latest membership and become operational.

\beginbsec{Recovery}
Hermes' fault tolerance properties enable a datastore to continue operating even 
in the presence of failures. 
However, as nodes fail, new nodes need to be added to the datastore to continue operating at peak performance. To add a new node, the membership is reliably updated, following which all other live replicas are notified of the new node's intention to join the replica group. Once all the replicas acknowledge this notification, the new node starts operating as a \textit{shadow replica} that participates as a follower for all 
of the 
writes but does not serve any client requests. Additionally, it reads chunks (multiple keys) from other replicas to fetch the latest values and reconstruct the datastore similarly to existing approaches~\cite{Dragojevic:2015,Ongaro:2011}. After reading the entire datastore, the shadow replica is up-to-date and transitions to operational state, whereby it is able to serve client requests.

\definecolor{myOrange}{RGB}{204, 102, 0}
\definecolor{myBlue}{RGB}{0, 114, 178}

\tikzstyle{sline} = [draw, -] 
\tikzstyle{inv} = [draw, -latex'] 
\tikzstyle{ack} = [draw, -latex', dashed] 
\tikzstyle{val} = [draw, -latex', dotted] 
\tikzstyle{client} = [rectangle, draw, fill=gray!30, 
    minimum width=1em, text centered,  minimum height=1em, node distance = 2em]
\tikzstyle{server} = [rectangle, draw, fill=blue!20, 
    minimum width=1em, text centered,  minimum height=1em, node distance = 1.5em]
\tikzstyle{rec} = [rectangle, draw=purple, text width=3.8em, text centered, minimum height=0.7em]
\tikzstyle{dashed_rectangle} = [rectangle, draw, dashed, 
    text width=3.8em, text centered, rounded corners, minimum height=1.4em]
\tikzstyle{session} = [draw, text width = 2.5em, rounded corners,node distance = 1em, yshift = -2em]

\tikzstyle{op} = [draw, |-|,line width=0.5mm]
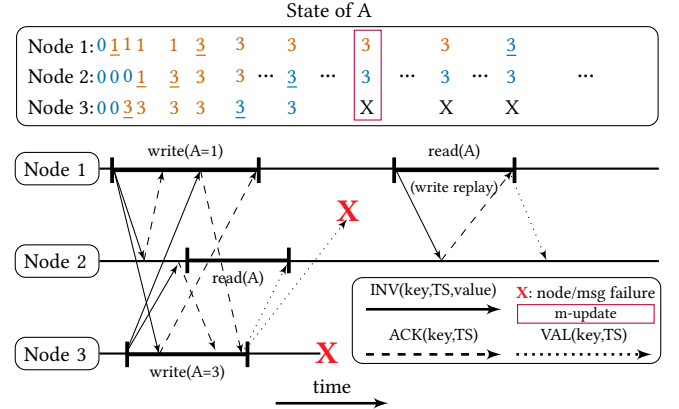
\begin{figure}[t]
\centering
\begin{tikzpicture}[auto]
    \node[session, xshift = - 4em, yshift = -0.5em] (node1) {\scalebox{0.8}{Node 1}};
    \node[session, yshift = -0.5em, below of = node1] (node2) {\scalebox{0.8}{Node 2}};
    \node[session, yshift = -0.5em, below of = node2] (node3) {\scalebox{0.8}{Node 3}};
    \path[draw, -latex',line width=0.1em] (1.5, -4) -- node[above] {\scalebox{0.8}{time}} ++ (1.5, 0);
    
    \node[rectangle, draw, text width=11em, minimum height=3.25em,
         rounded corners, xshift = 13em, yshift = -8.25em] (rect){};
    \path[inv, line width=0.1em] (2.7, -2.75) -- node[above] {\scalebox{0.65}{INV(key,TS,value)}} ++ (1.8, 0);
    \path[ack, line width=0.1em] (2.7, -3.35) -- node[above] {\scalebox{0.65}{ACK(key,TS)}} ++ (1.8, 0);
    \path[val, line width=0.1em] (4.7, -3.35) -- node[above] {\scalebox{0.65}{VAL(key,TS)}} ++ (1.8, 0);
    \draw[xshift = 16em, yshift = -7.25em]  node {\scalebox{0.65}{\textbf{\color{Red}X}: node/msg failure}};
    
    \draw[xshift = 16em, yshift = -8em]  node {\scalebox{0.6}{m-update}};
    \node[rec, xshift = 16em, yshift = -8em, text width=4.5em] (rm_rec){};
    \node[rec, minimum height=3.4em, xshift = 7.75em,
          yshift = 1.04em, text width=0.35em] (rm_rec_state){};
    
    \path[sline, thick]  (node1) -- node[left] {} ++ (8,0);
    \path[sline, thick]  (node2) -- node[left] {} ++ (8,0);
    \path[sline, thick]  (node3) -- node[left] {} ++ (3.5,0);
    
    \path[op] (-0.7, -0.9) -- node[above] {\scalebox{0.65}{write(A=1)}} ++ (2.0, 0);
    \path[op] (-0.5, -3.35) -- node[below] {\scalebox{0.65}{write(A=3)}} ++ (1.65, 0);

   \path[op] (0.3, -2.1) -- node[below] {\scalebox{0.65}{read(A)}} ++ (1.4, 0);
   \path[op] (3.05, -0.9) -- node[above] {\scalebox{0.65}{read(A)}} node[below] {\scalebox{0.6}{(write replay)}} ++ (1.65, 0); 
   
 \path[inv] (-0.65, -0.9) -- node[above] {} ++ (0.4, -1.2);
 \path[inv] (-0.65, -0.9) -- node[above] {} ++ (0.6, -2.45);
  \path[ack] (-0.25, -2.1) -- node[above] {} ++ (0.25, +1.2);
  \path[ack] (-0.05, -3.35) -- node[above] {} ++ (1.3, +2.45);
 \path[inv] (-0.5, -3.35) -- node[above] {} ++ (0.7, +1.2);
 \path[inv] (-0.5, -3.35) -- node[above] {} ++ (1, +2.45);
 \path[ack] (0.2, -2.1) -- node[above] {} ++ (0.5, -1.2);
 \path[ack] (0.5, -0.9) -- node[above] {} ++ (0.55, -2.45);
 \path[val] (1.05, -3.35) -- node[above] {} ++ (0.6, +1.2);
 \path[val] (1.05, -3.35) -- node[above] {} ++ (1.4, +1.8);
 
 \path[inv] (3.1, -0.9) -- node[above] {} ++ (0.6, -1.2);
 \path[ack] (3.7, -2.1) -- node[above] {} ++ (0.95, +1.2);
 
 \path[val] (4.6, -0.9) -- node[above] {} ++ (0.5, -1.2);
 
 \draw[xshift = 7em, yshift = -4.1em]  node {\color{Red}\scalebox{1.2}{\textbf{X}}};
 \draw[xshift = 6.2em, yshift = -9.5em]  node {\color{Red}\scalebox{1.2}{\textbf{X}}};

    \node[rectangle, draw, text width=23.6em, minimum height=3.75em,
         rounded corners, xshift = 6.6em, yshift = 1em] (rect){};
    \draw[xshift = 6.3em, yshift = 3.5em]  node {\scalebox{0.8}{State of A}};
    \draw[xshift = -3.85em, yshift = 2.15em]  node
    {\scalebox{0.8}{Node 1:}};
    \draw[xshift = -3.85em, yshift = 1em]  node {\scalebox{0.8}{Node 2:}}; 
    \draw[xshift = -3.85em, yshift = -0.15em]  node {\scalebox{0.8}{Node 3:}}; 
    \draw[xshift = 3.9em, yshift = 1em]  node {\scalebox{1}{...}};
    \draw[xshift = 6.25em, yshift = 1em]  node {\scalebox{1}{...}};
    \draw[xshift = 9.25em, yshift = 1em]  node {\scalebox{1}{...}};
    \draw[xshift = 12em, yshift = 1em]  node {\scalebox{1}{...}};
    \draw[xshift = 16em, yshift = 1em]  node {\scalebox{1}{...}};
    
    \draw[xshift = -2em, yshift = 1em]  node[text width = 1em]{
    \scalebox{0.8}{\color{myBlue}{0}} \\ 
    \scalebox{0.8}{\color{myBlue}{0}} \\ 
    \scalebox{0.8}{\color{myBlue}{0}}};
    \draw[xshift = -1.5em, yshift = 1em]  node[text width = 1em]{
    \scalebox{0.8}{\color{myOrange}{\underline{1}}} \\ 
    \scalebox{0.8}{\color{myBlue}{0}} \\ 
    \scalebox{0.8}{\color{myBlue}{0}}};
    \draw[xshift = -1em, yshift = 0.925em]  node[text width = 1em]{
    \scalebox{0.8}{\color{myOrange}{1}} \\ 
    \scalebox{0.8}{\color{myBlue}{0}} \\ 
    \scalebox{0.8}{\color{myOrange}{\underline{3}}}};
    \draw[xshift = -0.5em, yshift = 1em]  node[text width = 1em]{
    \scalebox{0.8}{\color{myOrange}{1}} \\ 
    \scalebox{0.8}{\color{myOrange}{\underline{1}}} \\ 
    \scalebox{0.8}{\color{myOrange}{3}}};
    \draw[xshift = 0.75em, yshift = 1em]  node[text width = 1em]{
    \scalebox{0.8}{\color{myOrange}{1}} \\ 
    \scalebox{0.8}{\color{myOrange}{\underline{3}}} \\ 
    \scalebox{0.8}{\color{myOrange}{3}}};
    \draw[xshift = 1.75em, yshift = 1em]  node[text width = 1em]{
    \scalebox{0.8}{\color{myOrange}{\underline{3}}} \\ 
    \scalebox{0.8}{\color{myOrange}{3}} \\ 
    \scalebox{0.8}{\color{myOrange}{3}}};
    \draw[xshift = 3.25em, yshift = 0.925em]  node[text width = 1em]{
    \scalebox{0.8}{\color{myOrange}{3}} \\ 
    \scalebox{0.8}{\color{myOrange}{3}} \\ 
    \scalebox{0.8}{\color{myBlue}{\underline{3}}}};
    \draw[xshift = 5.2em, yshift = 1em]  node[text width = 1em]{
    \scalebox{0.8}{\color{myOrange}{3}} \\ 
    \scalebox{0.8}{\color{myBlue}{\underline{3}}} \\ 
    \scalebox{0.8}{\color{myBlue}{3}}};
    \draw[xshift = 8.em, yshift = 1em]  node[text width = 1em]{
    \scalebox{0.8}{\color{myOrange}{3}} \\ 
    \scalebox{0.8}{\color{myBlue}{3}} \\ 
    \scalebox{0.8}{X}};
    \draw[xshift = 11em, yshift = 1em]  node[text width = 1em]{
    \scalebox{0.8}{\color{myOrange}{3}} \\ 
    \scalebox{0.8}{\color{myBlue}{3}} \\ 
    \scalebox{0.8}{X}};
    \draw[xshift = 13.5em, yshift = 1em]  node[text width = 1em]{
    \scalebox{0.8}{\color{myBlue}{\underline{3}}} \\ 
    \scalebox{0.8}{\color{myBlue}{3}} \\ 
    \scalebox{0.8}{X}};
\end{tikzpicture}
\vspace{-2.25em}
\caption{Concurrent writes to key A, then a read, followed by a node and a message failure which trigger a write replay on the last read. State of A shows the values of the replicas; blue represents Valid state, orange represents other states. Underlined values indicate a change in value and/or state.}
\vspace{-10pt}
\label{fig:Hermes-example}
\end{figure}
\subsection{Operational Example}
In this subsection, we discuss Figure~\ref{fig:Hermes-example}, which illustrates an example of Hermes' execution with reads and writes to key $A$. 
The purpose 
is to demonstrate the operation of Hermes while shedding light onto some of its corner cases in the presence of concurrency and failures. 
For simplicity, we assume no use of virtual node ids or any
latency optimizations (\S\ref{sec:optimizations}).

First, node 1 initiates a write ($A = 1$), by incrementing its local timestamp, broadcasting \CAP{INV} messages (solid lines) and transitioning key $A$ to Write state. Similarly, node 3 initiates another concurrent write ($A = 3$). 
Recall that \CAP{INV}s
in Hermes contain the key, the timestamp (including the 
\cid),
and the value to be written. We assume that key $A$ is initially stored with the same value (zero) and timestamp in all three nodes. 

Node 2 \CAP{ACK}s the \CAP{INV} message from node 1 (dashed line), updates its timestamp and value, and transitions key $A$ to Invalid state. Node 3 \CAP{ACK}s the \CAP{INV} of node 1, but it does not modify $A$ or its state since its local timestamp is higher (same version but higher 
\cid).
Subsequently, node 2 receives the \CAP{INV} from node 3, which has a bigger timestamp than the locally stored timestamp, resulting in an update in its local value and timestamp, all while remaining in Invalid state. 
Likewise, node 1 \CAP{ACK}s the \CAP{INV} of node 3, by updating the value, the timestamp, and transitioning to Trans state. 

Meanwhile, node 2 starts a read, but it is stalled since its local copy of $A$ is invalidated. 
Once node 3 receives all of the \CAP{ACK}s, it completes its own write by transitioning $A$ to the Valid state and broadcasts a \CAP{VAL} message (dotted lines) to the other replicas. 
When node 2 receives node's 3 \CAP{VAL} message, it transitions $A$ to Valid state and completes its stalled read. 

Once node 1 receives all of the \CAP{ACK}s it completes its write but transitions to the Invalid state.  This occurs because the write from node 3 took precedence over its own due to a higher timestamp, but the \CAP{VAL} from node 3 has not yet been received. Note that although the write from node 1 completes later than the concurrent write from node 3, it is linearized before the write of node 3 due to its lower timestamp 
(\cid).

As a last step, we consider a failure scenario, whereby the \CAP{VAL} message from node 3 to node 1 gets dropped and node 3 crashes. Thus, key $A$ in node 1 remains in the Invalid state.
Once leases expire and node 3 is detected as failed, the membership variable is reliably updated.
Subsequently, node 1 receives a read for $A$, but finds $A$ invalidated by a failed node.
Thus, node 1 triggers a write replay by broadcasting \CAP{INV} messages with the key's locally-stored timestamp and value (i.e., replaying node 3's original write). Crucially, the fact that \CAP{INV} messages contain both the timestamp and value to be written allows node 1 to safely replay node 3's write. Node 2 \CAP{ACK}s the \CAP{INV} from node 1 without applying it, since it already has the same timestamp. Once node 1 gets the \CAP{ACK} from node 2, it is able to unblock itself. Lastly, node 1 completes the replay of the write by broadcasting a \CAP{VAL} message to all of the live nodes (i.e., node 2, in this example).


\subsection{Read-Modify-Writes in Hermes}
\label{sec:hermes-RMWs}

So far, we have focused on read and write operations; however, Hermes also supports read-modify-write (\CAP{RMW}) atomics that are useful for synchronization (e.g., a compare-and-swap to acquire a lock).
In general, atomic execution of a read followed by a write to a key may fail if naively implemented with simple reads and writes. This is because a 
read followed by a write to a key is not guaranteed to be performed atomically since another concurrent write to the same key with a smaller logical timestamp could be linearized in-between the read-write pair, hence violating the \CAP{RMW} semantics.

For this reason, an \CAP{RMW} update in Hermes is executed similarly to a write, 
 but it is conflicting.
Hermes may abort an \CAP{RMW} which is
concurrently executed with another \textit{update} operation (either a write or another \CAP{RMW}) to the same key. Hermes commits an \CAP{RMW} if and only if the \CAP{RMW} has the highest timestamp amongst any concurrent updates to that key. Moreover, it purposefully assigns higher timestamps to writes compared to their concurrent \CAP{RMW}s. As a result, any write racing with an \CAP{RMW} to a given key is guaranteed to have a higher timestamp, thus safely aborting the \CAP{RMW}. Meanwhile, if only \CAP{RMW} updates are racing, the \CAP{RMW} with the highest node id will commit, and the rest will abort. 

More formally, Hermes always maintains safety and guarantees progress in the absence of faults
by ensuring two 
properties: (1) \textit{writes always commit}, and (2) \textit{at most one of possible concurrent \CAP{RMW}s to a key commits}. 
To maintain these properties, the following protocol alterations are needed:

\begin{itemize}[leftmargin=*]
\item\textbf{\scalebox{0.90}{Metadata}}: To distinguish between \CAP{RMW} and write updates, an additional binary flag (\VAR{RMW\_flag}) is included in \CAP{INV} messages. The flag is also stored in the per-key metadata to accommodate \textit{update replays}. 
\item\textbf{C\textsubscript{\CAP{TS}}}: When a coordinator issues an update, the version of the logical timestamp is incremented by one if the update is an \CAP{RMW} and by two if it is a write. 

\item\textbf{F\textsubscript{\CAP{RMW-ACK}}}: A follower \CAP{ACK}s an \CAP{INV} message for an 
\CAP{RMW} only if its timestamp is equal to or higher than the local one; otherwise, the follower responds with an \CAP{INV} based on its local state (i.e., same message used for write replay).

\item\textbf{C\textsubscript{\CAP{RMW}-abort}}: In contrast to non-conflicting writes, an \CAP{RMW} with pending \CAP{ACK}s is aborted if its coordinator receives an \CAP{INV}
to the same key 
with a higher timestamp.


\item
\textbf{C\textsubscript{\CAP{RMW}-replay}}: After an \CAP{RM} reconfiguration, the coordinator resets any gathered \CAP{ACK}s of a pending \CAP{RMW} and replays the \CAP{RMW} to ensure it is not conflicting.
\end{itemize}

\subsection{Summary}
This section introduced Hermes, a reliable membership-based protocol that guarantees linearizability. Hermes' decentralized broadcast-based design is engineered for high throughput and low latency. By leveraging invalidations and logical timestamps, Hermes enables 
efficient local reads and high-performance updates that are decentralized, fast, and inter-key concurrent. Writes (but not \CAP{RMW}s) in Hermes are also non-conflicting.
Finally, Hermes seamlessly recovers from a range of node and network faults thanks to
its write replays, enabled by {\em early value propagation} and logical timestamps.


\section{System}
\label{sec:system}

To evaluate the benefits and limitations of the Hermes protocol, we build HermesKV, an in-memory \CAP{RDMA}-based \CAP{KVS} with a typical read/write \CAP{API}. 
HermesKV is replicated across all the machines comprising a deployment and relies on the Hermes protocol to ensure the consistency of the deployment.
We choose \CAP{RDMA} networking to match the trend in modern datacenters towards offloaded network stacks and ultra-low latency fabrics instead of onloaded \CAP{UDP/TCP}~\cite{Marty:2019, Guo:2016}.

In \S\ref{sec:functional}, we present a functional overview of the HermesKV and briefly outline the implementation of its \CAP{KVS}. Subsequently, we describe \emph{Wings} (\S\ref{sec:wings}), our \CAP{RDMA}-based library which serves as the communication layer of the HermesKV.

\subsection{Overview and KVS} \label{sec:functional}
Each node in HermesKV is composed of a number of identical \emph{worker} threads. Each worker performs the following tasks: 1) decodes client requests; 2) accesses the local \CAP{KVS} replica; and 3) runs the Hermes protocol to complete requests. Client requests are distributed among the worker threads of the system. Requests can be either reads or writes. Worker threads communicate solely to coordinate writes (and write replays) as reads are completed locally. 

Our \CAP{KVS} is based on cc\CAP{KVS}~\cite{A&V:2018}, which is a version of \CAP{MICA}~\cite{Lim:2014} (found in~\cite{Kalia:2016}), but modified to support \CAP{CRCW} using seqlocks~\cite{Lameter:2005}.
Seqlocks are beneficial as they allow for efficient lock-free reads~\cite{Scott:2013}.
We further extend cc\CAP{KVS} to accommodate the Hermes-specific protocol actions, state transitions and request replies based on the replica state. 

The Hermes protocol is 
agnostic 
to the choice of a datastore and 
can be used with any datastore. We choose cc\CAP{KVS} since its minimalist design allows us to focus on the impact of the replication protocol itself without regard of idiosyncrasies or overheads of a commercial-grade datastore.

\subsection{Wings: an RDMA RPC layer for Hermes} \label{sec:wings}

State-of-the-art \CAP{RDMA}-based \CAP{KVS} designs such as \CAP{HERD}~\cite{Kalia:2014} and cc\CAP{KVS}~\cite{A&V:2018} have shown Remote Procedure Calls (\CAP{RPC}s) to be a highly effective design paradigm. Hence, we leverage 
\CAP{RDMA} Unreliable Datagram sends (\CAP{UD} sends) to build the \emph{Wings} library, a simple and efficient \CAP{RPC} layer over \CAP{RDMA}. Wings allows for opportunistic batching of multiple messages into one network packet, implements application-level flow control, provides support for broadcasts and enlists an array of \CAP{RDMA} low-level optimizations.

\beginbsec{Opportunistic Batching} 
The benefits of batching multiple application-level messages into a single network packet are well-known. Batching  amortizes the network header overhead, leading to better utilization of network bandwidth. 

Wings automatically performs opportunistic batching for all messages. The programmer provides Wings with a buffer that holds messages that need to be sent to various remote servers. 
Wings inspects the buffer in order to batch messages with the same receiver, then it creates a lightweight application-level header per batch specifying how many messages are batched and sends the packets. Note that the batching performed by Wings is \emph{opportunistic}, as it will never stall in order to form a batch; rather, Wings creates batches for the intended receivers only with readily available messages.

\beginbsec{Broadcast Primitive}
Wings implements software-based broadcasts as a series of unicasts to all members of a broadcast group.  Wings performs opportunistic batching for broadcasts in a similar manner as regular requests.

\beginbsec{Flow Control} 
Wings uses credit-based flow control~\cite{Kung:1994} to manage the data flow between the servers of a deployment. The programmer can specify whether the credit updates are \emph{explicit} or \emph{implicit}. Implicit credits are common in a communication pattern where a server sends a request and receives a response for that request; the response can be then treated as an implicit credit update. HermesKV leverages this feature when coordinating a write: the coordinator broadcasts invalidations to all remote replicas and treats the acknowledgments as credits updates. 
Explicit credits are needed for messages that do not require responses. HermesKV exploits explicit credits for the validation messages, as the protocol does not require validations to be acked. Instead, after receiving several validation messages, HermesKV nodes send explicit credit messages to the sender to inform it of their buffer availability. 
Similarly to
other Wings operations, explicit credits are opportunistically batched. The receiver polls a number of incoming messages and sends back a single explicit credit update message.

\beginbsec{RDMA Optimizations} 
In Wings, we build \CAP{RDMA} \CAP{RPC}s over \CAP{UD} sends following published low-level guidelines~\cite{Barak:2013, A&V:2018, Kalia:2016}. Transparent to the programmer, Wings amortizes and alleviates \CAP{PCI}e overheads. 
First, Wings performs doorbell batching and selective signaling when sending work requests to the \CAP{NIC}, and it inlines payloads inside the work requests when the payload is small enough (188B on our \CAP{NIC}) to reduce the required \CAP{NIC}-initiated \CAP{DMA}s per work request. 
Broadcasts are implemented as a linked list of work requests each with a different destination but all pointing to the same payload.
Moreover, explicit credit updates are header-only packets exploiting the \textit{immediate} header field~\cite{Barak:2015}. Thus, they are cheaper to transmit and due to the lack of a payload they reduce \CAP{PCI}e transactions on both sender and receiver sides.


\section{Experimental Methodology}
\label{sec:methodology}

\subsection{Evaluated Systems} 
\label{sec:meth-protocols}
We evaluate Hermes by comparing its performance with a majority-based and membership-based \CAP{RDMA}-enabled baseline protocols. To facilitate a fair protocol comparison, we study all protocols over a common multi-threaded KVS implementation based on HermesKV (as described in \S\ref{sec:system}). All protocols are implemented in C over the \CAP{RDMA} \emph{verbs} \CAP{API}~\cite{Barak:2015}.

\noindent The evaluated systems are as follows:
\squishlist
    \item \textbf{r\CAP{ZAB}} : In-house, multi-threaded, \CAP{RDMA}-enabled  
    \CAP{ZAB}~\cite{Reed:2008}.
    \item \textbf{r\CAP{CRAQ}}: In-house, multi-threaded, \CAP{RDMA}-based  \CAP{CRAQ}~\cite{Terrace:2009}.
    \item \textbf{HermesKV} : Implementation of Hermes as in \S\ref{sec:hermes} and \S\ref{sec:system}, without the latency optimization (O\textsubscript{3} from \S\ref{sec:optimizations}).
\squishend

Our evaluation mainly focuses on the comparison of HermesKV to r\CAP{ZAB} and r\CAP{CRAQ}, since they share the \CAP{KVS} and communication library, which allows us to isolate the effect of the protocol itself on performance. We also compare Hermes to \textbf{Derecho}~\cite{Jha:2019} (\S\ref{sec:eval-derecho}), the state-of-the-art \CAP{RDMA}-optimized open-source implementation of membership-based (i.e., virtually synchronous) Paxos.
Table~\ref{tab:eval} below summarizes the read and write features of the evaluated systems.
\begin{table}[h!]
\centering

\begin{adjustbox}{max width=0.47\textwidth}

\begin{tikzpicture}

\node (table) [inner sep=-0pt] {
\begin{tabular}{l|c|c|c|c|c}
\rowcolor[HTML]{9B9B9B} 
\cellcolor[HTML]{9B9B9B} & \multicolumn{2}{c|}{\cellcolor[HTML]{9B9B9B}\textbf{Local reads}} & \multicolumn{3}{c}{\cellcolor[HTML]{9B9B9B}\textbf{Writes}} \\ \cline{2-6} 
\rowcolor[HTML]{9B9B9B} 
\multirow{-2}{*}{\cellcolor[HTML]{9B9B9B}\textbf{System}} & \multicolumn{1}{c|}{\cellcolor[HTML]{9B9B9B}Leases} & \multicolumn{1}{c|}{\cellcolor[HTML]{9B9B9B}Consistency} & \multicolumn{1}{c|}{\cellcolor[HTML]{9B9B9B}Concurrency} & Latency (RTT) & \multicolumn{1}{l}{\cellcolor[HTML]{9B9B9B}Dec.} \\
\textbf{HermesKV} & one per RM & Lin & inter-key & 1 & \cmark \\
\textbf{rCRAQ} & one per RM & Lin & inter-key & O(\textit{n}) & \xmark \\
\textbf{rZAB} & none & SC & serializes all & 2 $\dagger$ & \xmark \\
\textbf{Derecho} & none & SC & serializes all & 1 $\ddagger$ & \cmark
\end{tabular}
};
\draw [rounded corners=.3em] (table.north west) rectangle (table.south east);

\end{tikzpicture}

\end{adjustbox}

\caption{
Comparison of read and write features for the evaluated systems. 
SC: sequentially consistent; RM: reliable membership; Dec: decentralized; \textit{n}: number of replicas; $\dagger$1 RTT for master's writes; $\ddagger$lock-step commit.}
\label{tab:eval}
\vspace{-25pt}
\end{table}

\vspace{-3pt}
\subsubsection{rZAB}
\vspace{-2pt}
\label{sec:ZAB}
In \CAP{ZAB} protocol, one node is the leader and the rest are followers. A client can issue a write to any node, which in turn propagates the write to the leader. 
The leader receives writes from all nodes, serializes them and proposes them by broadcasting atomically to all followers. The followers send back acknowledgements (\CAP{ACK}s) to the leader; on receiving a majority of \CAP{ACK}s for a given write, the leader commits the write locally and broadcasts commits to the followers.

A client's read can be served locally by any node without any communication as long as the last write of that client has been applied in that node. 
%
However, local reads in \CAP{ZAB} are sequentially consistent (\CAP{SC}), which is weaker than Lin. Problematically, the fact that \CAP{ZAB} is not Lin leads to a performance issue on writes.
This is because, in contrast to the stricter Lin, sequential consistency (\CAP{SC}) is not compositional~\cite{Attiya:1994}. As a result, it is not possible to deploy independent instances (e.g., per-key) of SC protocols such as \CAP{ZAB} to increase the concurrency of writes because the composition of those instances would violate \CAP{SC}.
If a  \CAP{ZAB} client requires linearizable reads, then it can issue a \emph{sync} command prior to the read. A sync is completed similarly to a write, necessarily increasing the read latency. We do not evaluate linearizable reads, to get the upper bound performance of the \CAP{ZAB} protocol.

\beginbsec{rZAB optimizations}
We apply to r\CAP{ZAB} all HermesKV optimizations and utilize the \CAP{RDMA} Multicast~\cite{Barak:2015} to tolerate \CAP{ZAB}'s asymmetric (i.e. leader-oriented) network traffic pattern.
Our highly optimized, \CAP{RDMA} implementation of \CAP{ZAB} outperforms the open-source implementation of Zook\-eep\-er (evaluated in ~\cite{Jin:2018}) by three orders of magnitude. Of course, Zookeeper is a production system incorporating features beyond the \CAP{ZAB} protocol, such as client tracking and check-pointing to disk. By evaluating a lean and optimized version of just \CAP{ZAB}, we are 
facilitating a fair protocol comparison.

\subsubsection{rCRAQ}
\CAP{CRAQ} 
affords local reads and inter-key concurrent, but not decentralized, writes (\S\ref{sec:craq} details the \CAP{CRAQ} protocol). We identify two undesirable properties of \CAP{CRAQ}: 1) writes must traverse multiple hops before completing, adversely affecting the system's latency; and 2) the nodes of the chain are generally not well balanced, 
in terms of the amount of work performed per-packet potentially affecting the system's throughput. 
To evaluate how these properties affect performance, we study our own \CAP{RDMA}-enable version of \CAP{CRAQ} (r\CAP{CRAQ}), that enjoys all optimizations available in HermesKV.

\vspace{-2pt}
\subsection{Testbed}
\vspace{-3pt}
We conduct our experiments on a cluster of 7 servers interconnected via a 12-port Infiniband switch (Mellanox \CAP{MX6012F}). Each machine runs Ubuntu 18.04 and is equipped with two 10-core \CAP{CPU}s (Intel Xeon E5-2630v4) with 64 \CAP{GB} of system memory and a single-port 56Gb Infiniband \CAP{NIC} (Mellanox \CAP{MCX455A-FCAT PCI3} x16). Each \CAP{CPU} has 25 \CAP{MB} of L3 cache and two hardware threads per core.  
We disable turbo-boost, pin threads to cores and use huge pages (2 \CAP{MB}) for the \CAP{KVS}.
The \CAP{KVS} consists of one million key-value pairs, replicated in all nodes. Unless stated otherwise, we use keys and values of $8$ and $32$ bytes, respectively; which are accessed uniformly.

\section{Evaluation}
\label{sec:evaluation}

\begin{figure*}[t]
  \hspace{30pt}
  \begin{subfigure}{0.32\textwidth}
  \centering
  \includegraphics[width=0.9\linewidth]{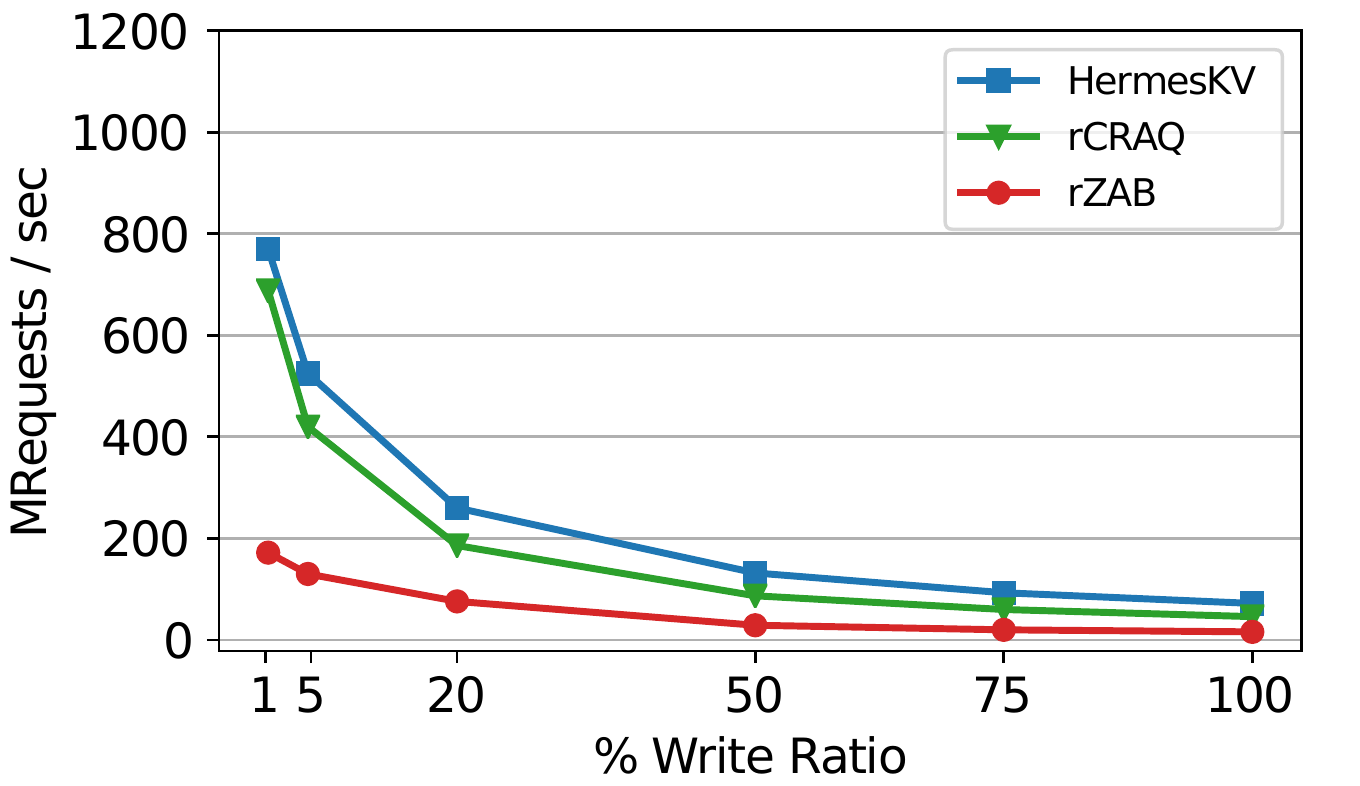}
  \vspace{-5pt}
  \subcaption{Uniform}
  \label{fig:write-rate}
  \end{subfigure}%
\hspace{100pt}
  \begin{subfigure}{0.32\textwidth}
  \centering
  \includegraphics[width=0.9\linewidth]{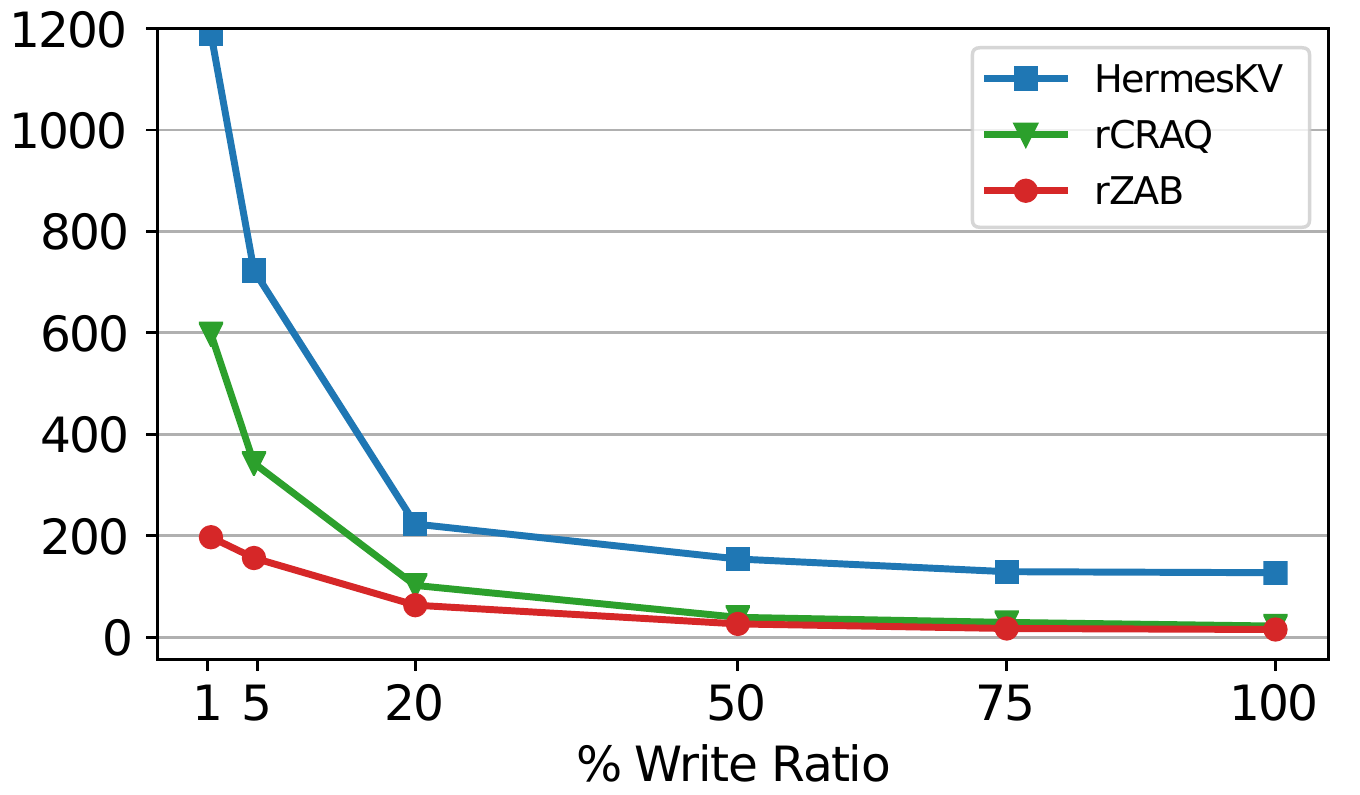}
  \vspace{-5pt}
  \subcaption{Skewed [Zipfian exponent 0.99]}
  \label{fig:skew-line}
  \end{subfigure}
  \vspace{-10pt}
  \caption{Throughput for $1\%$ to $100\%$ write ratio. [5 nodes]}
  \vspace{-13pt}
\end{figure*}

\subsection{Throughput on Uniform Traffic} 
\label{sec:throughput} 

Figure~\ref{fig:write-rate} shows the performance of HermesKV, r\CAP{CRAQ} and r\CAP{ZAB} while varying the write ratio under uniform traffic.\goodbreak\noindent 

\beginbseceval{Read-only} For read-only, all three systems exhibit identical behaviour, achieving $985$ Million Requests per second (MReqs/s), as all systems perform reads locally from all replicas. To reduce clutter we omit the read-only from the figure.

\beginbseceval{HermesKV} 
At a $1\%$ write ratio (Figure~\ref{fig:write-rate}), HermesKV achieves $770$ MReqs/s, outperforming both baselines ($12\%$ better than r\CAP{CRAQ} and $4.5\times$ better than r\CAP{ZAB}). 
As the write ratio increases, the throughput of HermesKV gradually drops, reaching $72$ MReqs/s on a write-only workload. The throughput degradation at higher write ratios is expected because writes require an exchange of messages over the network, which cost both \CAP{CPU} cycles and network bandwidth.

At 20\% write ratio, HermesKV significantly outperforms the baselines ($40\%$ over r\CAP{CRAQ}, $3.4 \times$ over r\CAP{ZAB}). The reason for HermesKV's good performance compared to alternatives is that it combines local reads with high-performance writes. 

\beginbseceval{rCRAQ} 
The \CAP{CRAQ} protocol is well-suited for high throughput, comprising both inter-key concurrent writes and local reads. Nevertheless, r\CAP{CRAQ} performs worse than HermesKV across all write ratios, with the gap widening as write ratios increase. That difference has its root in the design of \CAP{CRAQ}.

Firstly, reads in \CAP{CRAQ} are not always local: if a non-tail node is attempting to serve a read for a key for which it has seen a write but not an \CAP{ACK}, then the tail must be queried to find out whether the write has been applied or not. Therefore, increasing the write ratio has an adverse effect on the reads, as more reads need to be served remotely via the tail node.

This disadvantage hints to a more important design flaw: the \CAP{CRAQ} design is heterogeneous, mandating that nodes assume one of three different roles -- head, tail or intermediate -- where each role has different responsibilities. As such, 
load is not equally
balanced, so the system is always bottlenecked by the node with the heaviest responsibilities. For instance, at high ratios, the tail node is heavily loaded as it receives read queries from all nodes. Meanwhile, at low write ratios, the tail has fewer responsibilities than an intermediate node, as it only propagates acknowledgements up the chain, whilst an intermediate must also propagate writes downstream. 

\beginbseceval{rZAB} 
As expected, \CAP{ZAB} fails to achieve high throughput at non-zero write ratios as it imposes a strict ordering constraint on {\em all} writes at the leader. The strict ordering makes it difficult to extract concurrency, inevitably causing queuing of writes and delaying the subsequent reads within each session. At $1\%$ write ratio, r\CAP{ZAB} achieves $172$ MReqs/s, which drops to a mere $16$ MReqs/s for a write-only workload. 

\subsection{Throughput under Skew} 
\label{sec:skew}
We next explore how the evaluated protocols perform under access skew. We study an access pattern that follows a power-law distribution with a Zipfian exponent of 0.99, as in \CAP{YCSB}~\cite{Cooper:2010} and recent studies~\cite{Dragojevic:2014, A&V:2018, RNovakovic:2016}. Figure~\ref{fig:skew-line} shows the performance of the three protocols when varying the write ratio from $1\%$ to $100\%$. We discuss read-only separately.

\beginbseceval{Read-only} 
Similarly to the uniform read-only setting, all three protocols achieve identical performance ($4183$  MReq/s) due to their all-local accesses. Unsurprisingly, the read-only performance under the skewed workload is higher than the uniform performance for all protocols. This is because under a skewed workload there is temporal locality among the popular objects, which is captured by the hardware caches.

\beginbseceval{HermesKV} 
HermesKV gracefully tolerates skewed access patterns, especially at low write ratios (achieving $1190$ MReq/s at $1\%$ write ratio). Repeatedly accessing popular objects cannot adversely affect HermesKV write throughput, as concurrent writes to the popular objects can proceed without stalling (as explained in \S~\ref{sec:conc-writes}). Meanwhile, read throughput thrives under a skewed workload as reads are always local in HermesKV, and as such can benefit from temporal locality.

\beginbseceval{rCRAQ} 
Similarly, r\CAP{CRAQ} benefits from temporal locality when accessing the local \CAP{KVS}, while write throughput is unaffected by the skew,
as multiple writes for the same key can concurrently flow through the chain. The problem, however, is that non-tail nodes cannot complete reads locally if they have seen a write for the same key but have not yet received an \CAP{ACK}. In that case, the tail must be queried. Under skew,
such cases become frequent, with reads to popular objects often serviced by the tail and not locally. Thus, at higher write ratios, the tail limits r\CAP{CRAQ}'s performance. 

\beginbseceval{rZAB} 
r\CAP{ZAB} is not affected by the conflicts created by the skewed access pattern, as it already serializes all writes irrespective of the object they write. In practice, r\CAP{ZAB} performs slightly better under skew as hardware caches are more effective due to better temporal locality for popular objects.

\begin{figure*}[t]
  \begin{subfigure}[b]{0.33\textwidth}
    \includegraphics[width=\textwidth]{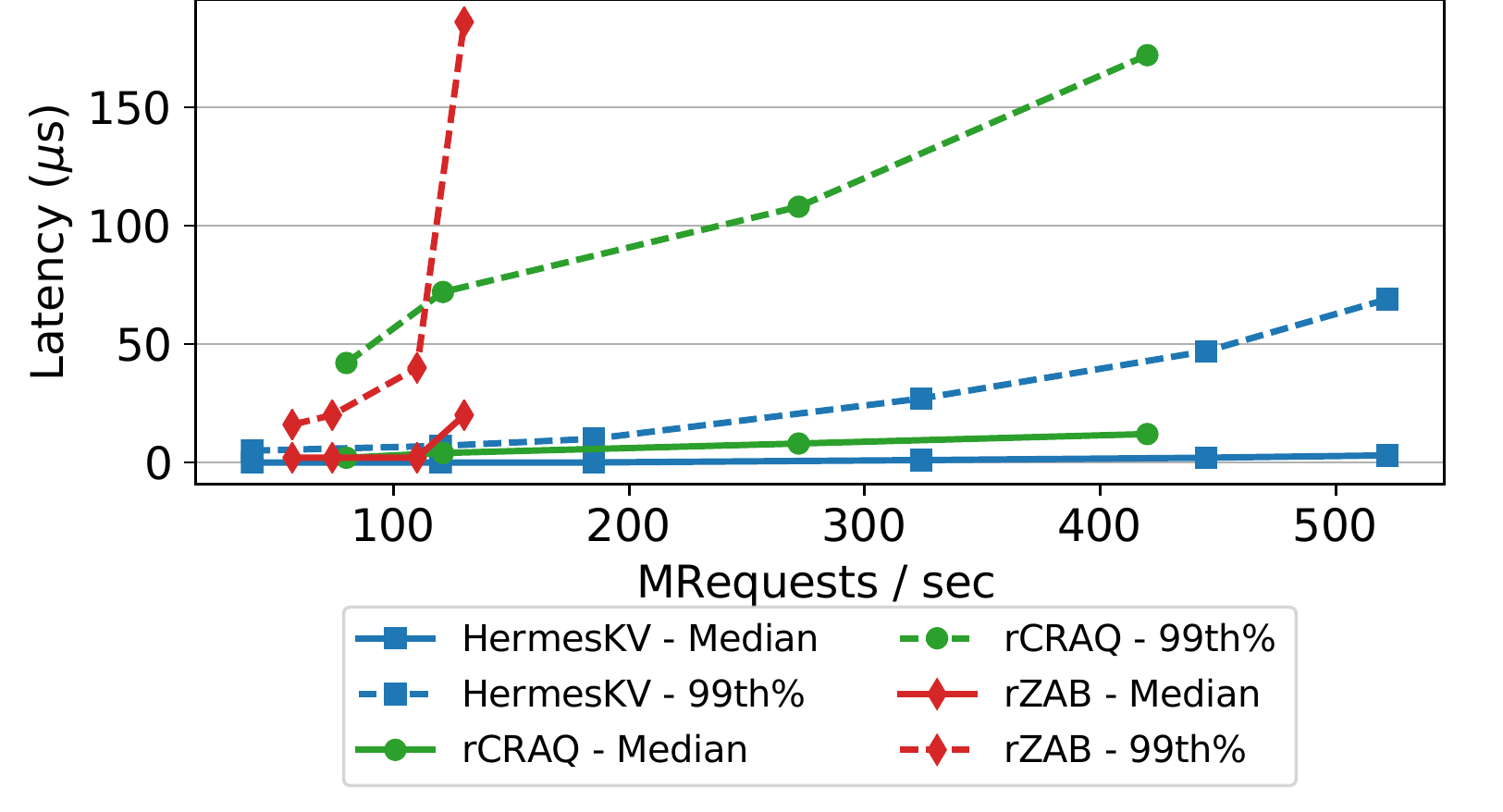}
    \captionsetup{width=0.85\linewidth}
    \vspace{-1.8em}
    \caption{Latency vs throughput. [Uniform traffic, 5\% write ratio]}
    \label{fig:latency}
  \end{subfigure}
  \begin{subfigure}[b]{0.33\textwidth}
    \includegraphics[width=\textwidth]{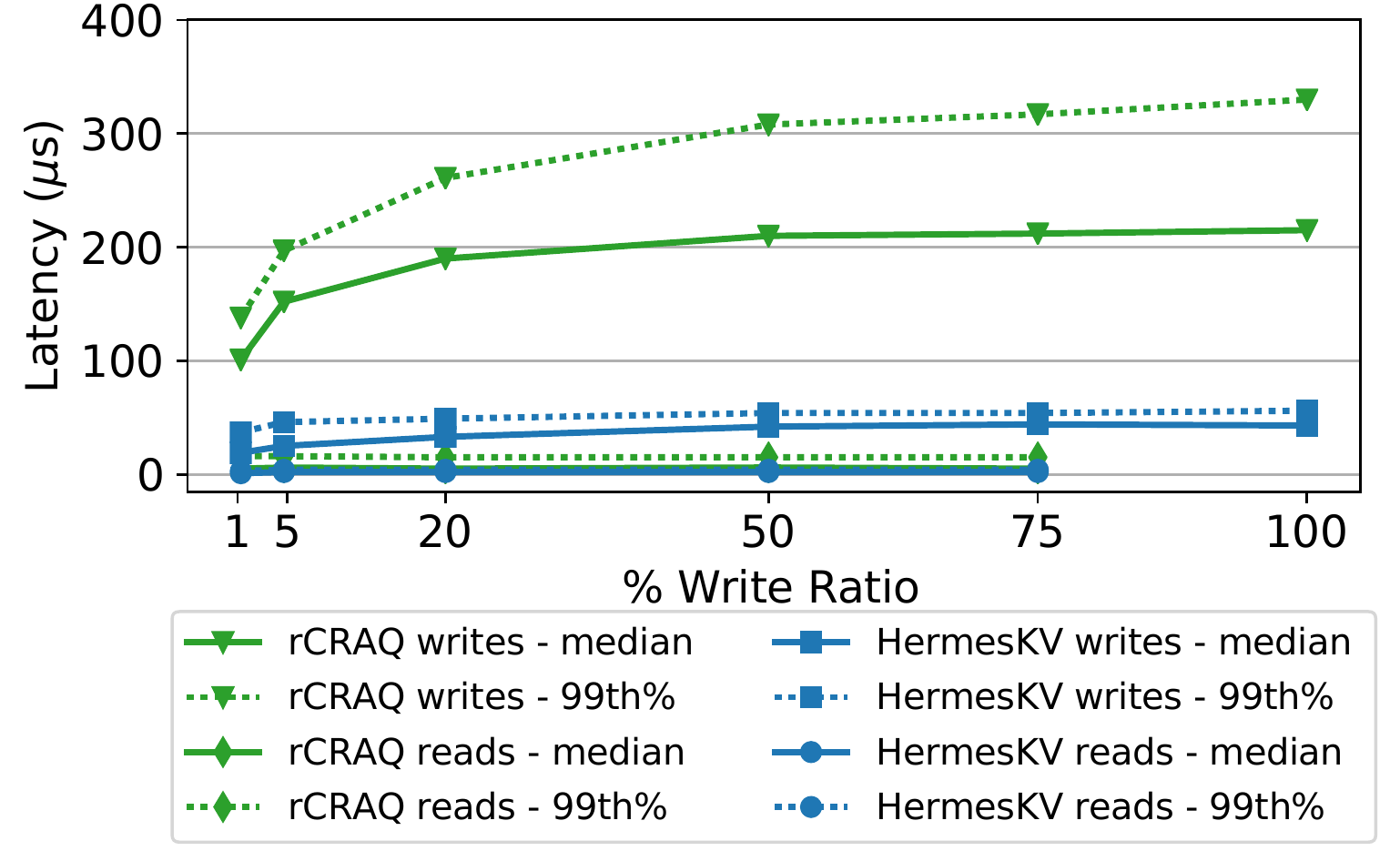}
    \captionsetup{width=0.85\linewidth}
    \vspace{-1.8em}
    \caption{Median and 99th\% [Uniform traffic]}
    \label{fig:latency-uni}
  \end{subfigure}
  \begin{subfigure}[b]{0.33\textwidth}
    \includegraphics[width=\textwidth]{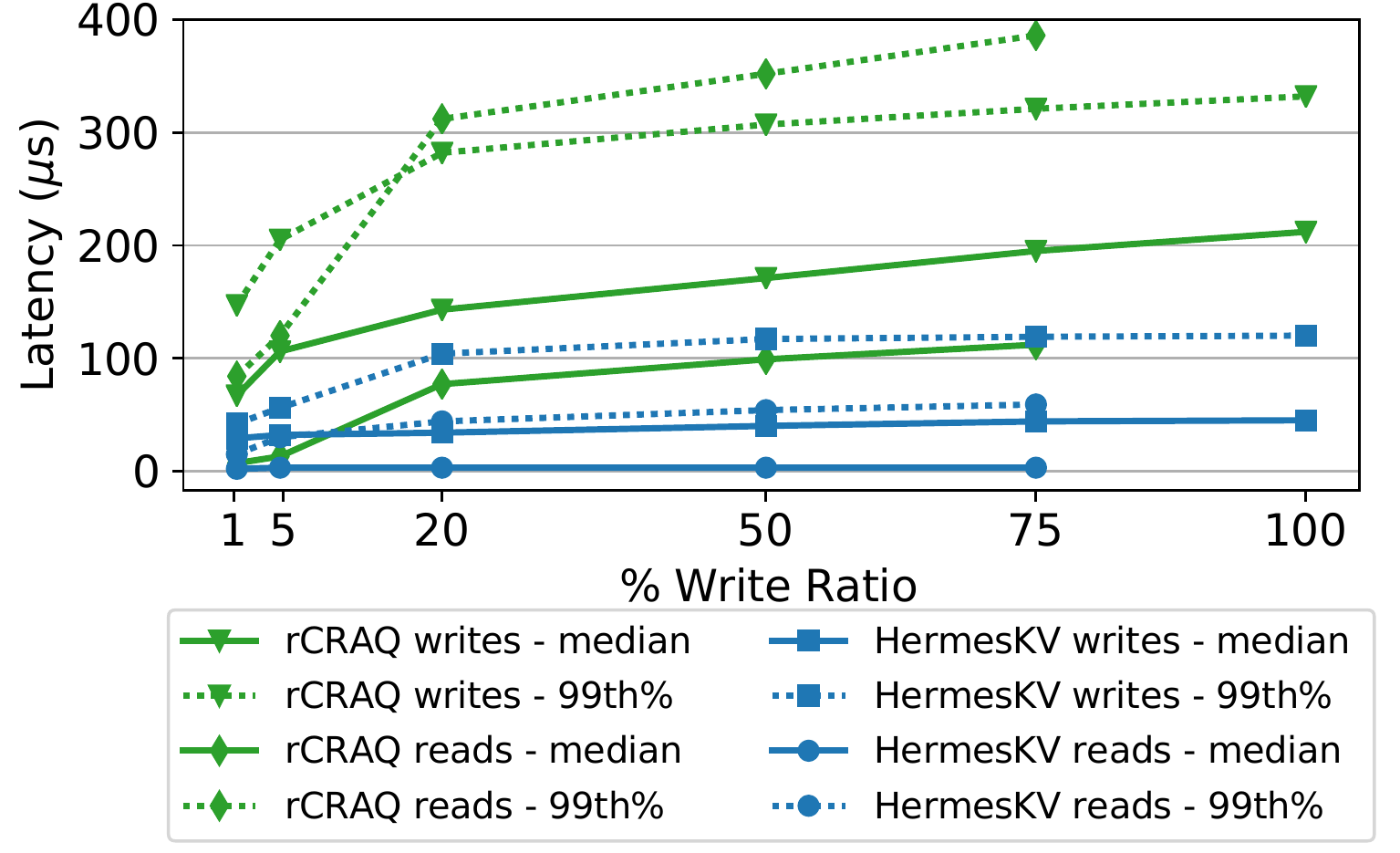}
    \captionsetup{width=0.85\linewidth}
    \vspace{-1.8em}
    \caption{Median and 99th\% [Zipfian 0.99]}
    \label{fig:latency-skew}
  \end{subfigure}
  \vspace{-2.5em}
  \caption{Latency analysis. [5 nodes] }
  \vspace{-10pt}
\end{figure*}
\vspace{-5pt}
\subsection{Latency Analysis}
\label{sec:latency}
\subsubsection{Latency vs Throughput}
Figure~\ref{fig:latency} illustrates the median (50th\%) and the tail (99th\%) latencies of the three protocols as a function of their throughput at $5\%$ write ratio. We measure latency of each request from the beginning of its execution to its completion. 

All three systems execute reads locally, while writes incur protocol actions that include traversing the network. Therefore, at 5\% write ratio, we expect the median latency of all protocols to be close to the latency of a read and the tail latency to be that of a write. Consequently, the gap between the median and the tail latency is to be expected for all systems and should not be interpreted as unpredictability.

\beginbseceval{HermesKV} 
The median latency of HermesKV is the latency of a read, and as expected, is consistently very low (on the order of $1\mu s$) even at peak throughput because reads are local. The tail latency is determined by the writes. The tail latency increases with the load, because writes traverse the network and thus can be subject to queuing delays as load increases. At peak throughput, the tail latency of HermesKV is $69\mu s$.

\beginbseceval{rCRAQ} 
In r\CAP{CRAQ} the median latency is the latency of a read, and as such, is typically on the order of a few microseconds. As expected, the tail latency, which corresponds to a write, is consistently high -- at least $3.6\times$ larger than HermesKV at the same throughput points -- ranging from $42\mu s$ at lowest load to $172\mu s$ at peak load. The high write latency is directly attributed to the protocol design as writes in r\CAP{CRAQ} need to traverse multiple network hops, incurring both the inherent network latency and the queuing delays in all the nodes.

\beginbseceval{rZAB} 
As the other two protocols, r\CAP{ZAB} achieves a low median latency because of its local reads, but even at 
moderate throughput,
its tail latency is much larger
(e.g., more than 3.6$\times$ than that of Hermes at 75MReq/s)
because of the high latency of the writes that must serialize on the leader. 

\subsubsection{Latency vs Write ratio} 

Figures~\ref{fig:latency-uni} and~\ref{fig:latency-skew} depict the median and tail latencies of reads and writes separately, under both skewed and uniform workloads, when operating at peak throughput of \CAP{CRAQ}~--~which corresponds roughly to 50-85\% of HermesKV peak throughput. r\CAP{ZAB} cannot achieve high enough throughput to be included in the figures. 

\beginbseceval{Uniform} 
HermesKV delivers very low, tightly distributed latencies across all write ratios, for both reads ($2\mu s$-$15\mu s$) and writes ($29\mu s$-$42\mu s$). As expected, r\CAP{CRAQ} exhibits a similar behaviour for reads but not for writes. r\CAP{CRAQ} write latencies are at least $3.9 \times$ to $5.9\times$ larger than the corresponding write latencies of HermesKV, with median latencies ranging from $101$ to $215\mu s$ while the tail latencies range from $138$ to $330\mu s$.

\beginbseceval{Skew} 
Under skew the tail latencies of both reads and writes increase in HermesKV, because reads and writes are more likely to conflict on popular objects. The tail read latency is the latency of a read that stalls waiting for a write to return; not surprisingly that latency is roughly equal to the median latency of a write. 
Similarly,
the tail latency of a HermesKV write increases up to $120\mu s$  because in the worst case without failures a write might need to wait an already outstanding write (to the same key) issued from the same node.

In r\CAP{CRAQ}, the latencies of writes remain largely unaffected, compared to the uniform workload. However, the behaviour of reads changes radically because reads are far more likely to conflict with writes under skew; such reads are sent to the tail node. Consequently, the tail node becomes very loaded, which is reflected in both the median (up to $112\mu s$) and tail (up to $386\mu s$) read latencies. This is a very important result; while high write latencies are expected of r\CAP{CRAQ}, we show that reads latencies can suffer as well, making \CAP{CRAQ} an undesirable protocol for systems that target low latency.

\begin{figure*}[t]
\setkeys{Gin}{width=1.05\linewidth}
\begin{tabularx}{\linewidth}{XXX}
\includegraphics{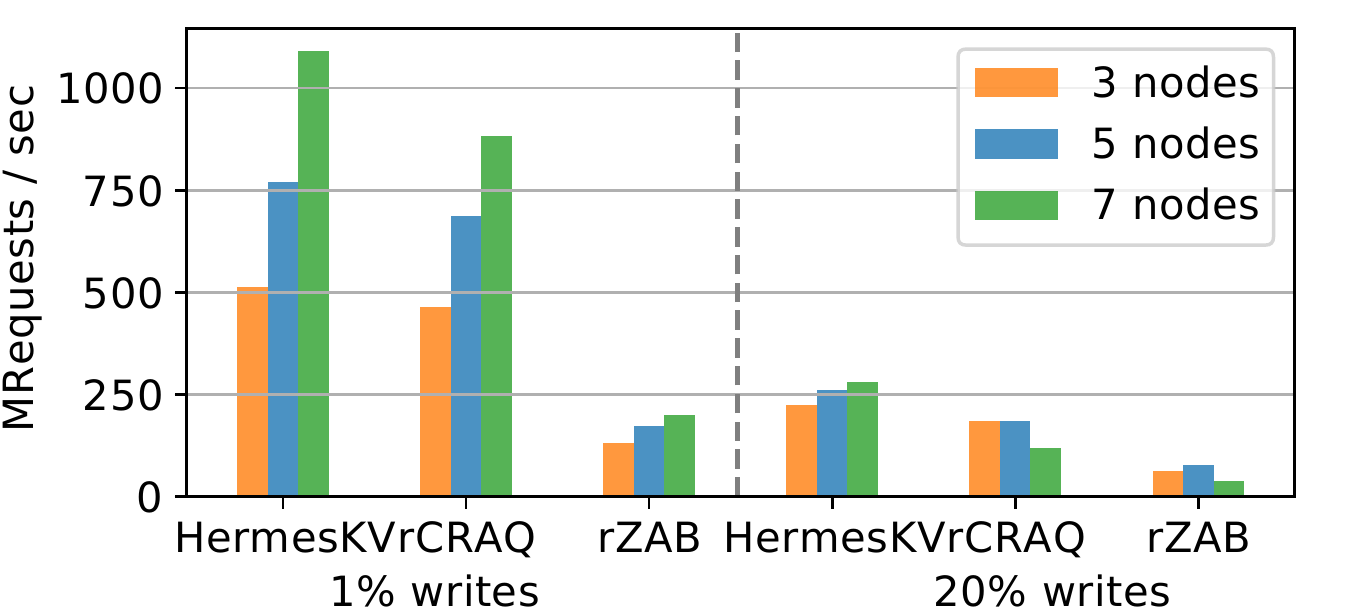}
\vspace{-2.25em}
\caption{Scalability study. [Uniform traffic] }
\label{fig:scalability-line}
&
\includegraphics{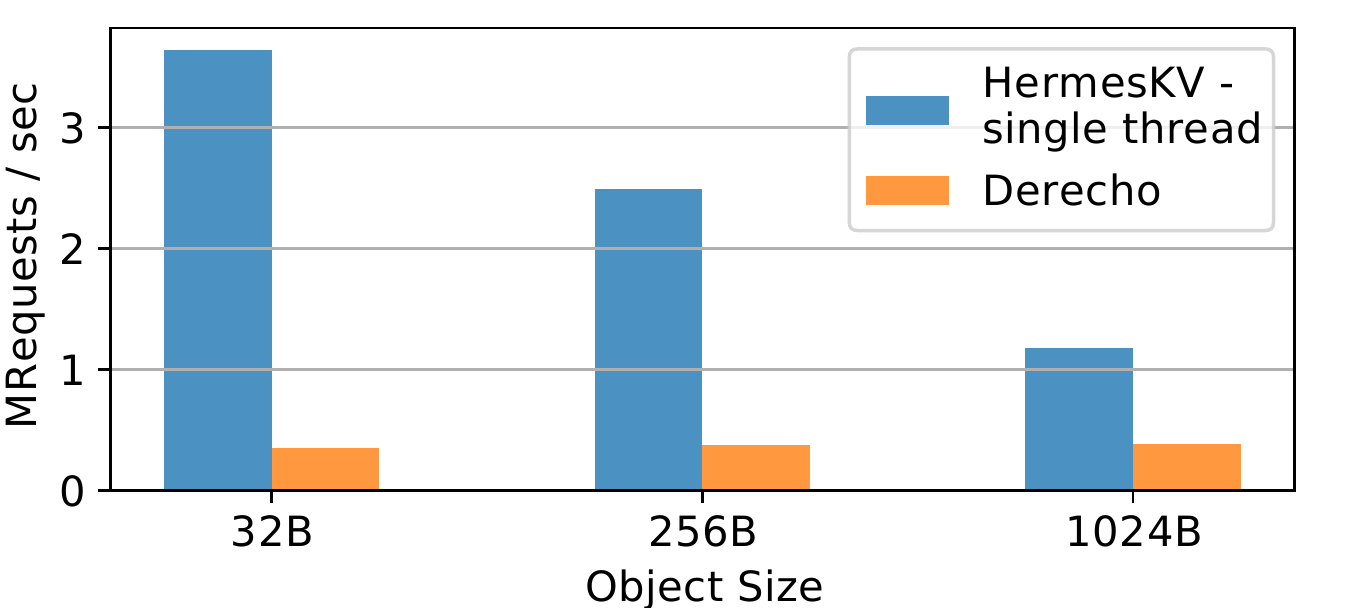}
\vspace{-2.25em}
\caption{Comparison to Derecho. [Uniform traffic, 5 nodes, write-only]}
\label{fig:derecho}
&
\input{3_misc/Eval-failure.tex}
\vspace{-2.25em}
\caption{HermesKV under failure. [Uniform traffic, 5 nodes, timeout=150ms]}
\label{fig:failure-throughput}
\vspace{-1.5em}
\end{tabularx}
\vspace{-20pt}
\end{figure*}

\subsection{Scalability Study}
\label{sec:scalability}

To investigate the scalability of the evaluated protocols, we measure their performance by varying the replication degree. Figure~\ref{fig:scalability-line} depicts the throughput of the three protocols under write ratios of $1\%$ and $20\%$  for 3, 5 and 7 machines.

\beginbseceval{HermesKV} 
Reads in HermesKV are always local and thus their overhead is independent of the number of replicas, allowing HermesKV to take advantage of the added replicas to increase its throughput. Therefore, HermesKV's scalability is dependent on the write ratio, achieving almost linear scalability with the number of replicas at $1\%$ writes, while maintaining its performance advantage at 20\% write ratio.

\beginbseceval{rCRAQ} 
When scaling r\CAP{CRAQ}, the expectations are similar to HermesKV: reads are scalable, but writes are not. However, scaling the 
replicas in \CAP{CRAQ} implies extending the size of the chain. Consequently, more non-tail nodes redirect their reads to the tail node. Thus, the tail becomes loaded, degrading read throughput, while also creating back-pressure in the chain which adversely affects write throughput. That phenomenon is apparent in Figure~\ref{fig:scalability-line}; at $20\%$ write ratio, r\CAP{CRAQ} throughput degrades when the chain is extended from 5 to 7 nodes. 

\beginbseceval{rZAB} 
r\CAP{ZAB} also performs reads locally, and thus is expected to see a benefit from greater degrees of replication at low write ratios. However, write requests incur a large penalty in r\CAP{ZAB}, as the leader receives and serializes writes from all machines. When the leader cannot keep up with the write stream, the replicas inevitably fall behind as the reads stall waiting for the writes to complete, and the writes are queued on the leader. Indeed, in Figure~\ref{fig:scalability-line}, we observe that even though r\CAP{ZAB} scales well for a read-dominant workload, at a $20\%$ write ratio, increasing the replication degree from 5 to 7 cuts the performance almost in half. Our results are in line with the original scalability analysis of Zookeeper~\cite{Hunt:2010}.

\subsection{Comparison to Derecho}
\label{sec:eval-derecho}
In this section, we compare HermesKV throughput
with the \CAP{RDMA}-optimized open-source Derecho~\cite{Jha:2019}, the state-of-the-art
mem\-ber\-ship-based 
variant of
Paxos. 
Derecho's codebase partitions work at each node across several threads (3-4), but does not support higher degrees of threading. 
For a fairest possible comparison, we limit HermesKV to a single thread. 

Figure~\ref{fig:derecho} shows throughput of a write-only workload, while varying the object size from \CAP{32B} to \CAP{1KB} 
-- such relatively small object sizes are typical for datastore workloads~\cite{Atikoglu:2012,Lim:2014}.
Although HermesKV is constrained to a single thread, it outperforms Derecho by an order of magnitude on small object sizes (\CAP{32B}), while maintaining its benefit even on larger objects (\CAP{$3\times$} at \CAP{1KB}). 
Derecho increases the performance of its totally ordered writes by exploiting monotonic predicates~\cite{Jha:2019}. Nevertheless, due to its lock-step delivery and its inability to offer inter-key concurrent writes, it fails to match the performance of Hermes.
We note that HermesKV's throughput naturally decreases as the object size increases and more bytes per request are transferred.

\vspace{-1pt}
\subsection{Throughput with Failures} 
\vspace{-1pt}
\label{sec:failure-study}
In order to study the behaviour of HermesKV when a failure occurs, we implement \CAP{RM}
in a similar manner with~\cite{Kakivaya:2018} and integrate it with HermesKV. Figure~\ref{fig:failure-throughput} depicts the behaviour of HermesKV when a failure is injected at 1, 5, and 20\% write ratios in a five node deployment and a conservative timeout of 150ms. The throughput drops to zero almost immediately after the failure, because all live nodes are blocked waiting for acknowledgements from the failed node. After the timeout expires, the machines reach agreement
(via a majority-based protocol) to reliably
remove the failed node from the membership, and subsequently continue operating with four nodes. The agreement part of the protocol entails exchanging a handful of small messages over an unloaded \CAP{RDMA} network, which takes just a few microseconds and is not noticeable in the figure. 
The recovered, steady-state throughput is lower after the failure, because one node is removed from the replica group.

\section{Related Work}
\vspace{-4pt}
\label{sec:related-work}

\beginbsec{Consensus and atomic broadcast}
State machine replication (\CAP{SMR})~\cite{Schneider:1990}
provides linearizability by explicitly ordering all client requests (reads and writes), and requiring 
all replicas to execute the requests in the determined order. \CAP{SMR} can be implemented using any fault-tolerant consensus or atomic broadcast algorithm to order the requests.  
Numerous such algorithms have been proposed~\cite{Chandra:1996, Birman:1987, Liskov:2012, Oki:1988}, the most popular being variants of Paxos~\cite{Lamport:1998}.
Recent works present optimized variants of these protocols that exploit commutative operations~\cite{Moraru:2013, Aguilera:2000, Lamport:2005} and rotating coordinators~\cite{Mao:2008}. 
Others leverage a ring-based topology~\cite{Marandi:2010, Amir:1995, Guerraoui:2007}, similarly to \CAP{CRAQ}, to increase throughput but at the cost of latency.


Most of these
protocols are majority-based and sacrifice performance for a failure model without \CAP{RM} support.
Therefore, they typically enforce strong consistency at the cost of performance, by sacrificing either local reads or concurrency.
An abundance of such protocols forfeits local reads~\cite{Lamport:1998, Moraru:2013, Attiya:1995, Lamport:2001, Lamport:2006, Lamport:2005, Mao:2008, Ekstrom:2016, Birman:1987, Bolosky:2011, Ongaro:2014, Li:2016-NoPaxos, Marandi:2011, Poke:2015}, thus incurring a significant 
penalty on read-dominant datastore workloads. 

Meanwhile, protocols that allow local reads sacrifice performance on writes. A recent atomic broadcast protocol offering local reads does so by relaxing consistency and applying writes in lock-step~\cite{Poke:2017}. Chandra et al.~\cite{Chandra:2016}, present a protocol with linearizable local reads 
through object leases, which serializes writes on a leader.
\CAP{ZAB}~\cite{Reed:2008}, a characteristic example of such protocols, enables local reads and serializes writes on a leader but without using object leases, thus increasing performance but at the cost of consistency. 
As shown in our evaluation, Hermes significantly outperforms \CAP{ZAB} with its decentralized and inter-key concurrent writes.

\beginbsec{Per-key leases} 
Linearizable protocols that use object leases for local reads, such as~\cite{Chandra:2016, Baker:2011, Moraru:2014}, could be deployed per-key (\ie one protocol instance for each key) to match the inter-key concurrency, but not latency,
of writes in Hermes. However, this mandates a lease for \textit{each} individual key, which is not scalable for realistic datastores with millions of keys. In this approach, for linearizable local reads, leases must be continuously renewed for each key --- even in the absence of writes or reads. This renewal costs at least $\Theta(n)$ messages (\textit{n} = number of replicas) per key and must occur 
before each lease expires,
causing significant network traffic. Moreover, the lease duration cannot be made very long
since this would translate into similarly long unavailability upon a fault. In contrast, Hermes, with its invalidating writes and just a single \CAP{RM} lease per replica, offers local reads while being fully inter-key concurrent at a message cost independent of the number of the keys stored by the datastore.  

\beginbsec{Hardware-assisted replication}
Some proposals leverage hardware support to reduce the latency of reliable replication, such as \CAP{FPGA} offloading~\cite{Istvan:2016} and programmable switches~\cite{Jin:2018, Dang:2015, Li:2016-NoPaxos}. For instance, 
Zhu et al.~\cite{zhu:2019} use programmable switches for in-network conflict detection, to allow local reads from any replica. 
Other works tailor reliable protocols by exploiting \CAP{RDMA}~\cite{Poke:2015, Wang:2017, Behrens:2016}. 
Hermes offers local reads without hardware support. When evaluated over 
\CAP{RDMA}, Hermes significantly outperforms Derecho, which represents the state-of-the-art of \CAP{RDMA}-based approaches (\S\ref{sec:eval-derecho}).

\beginbsec{Optimized reliable replication}
A recent work~\cite{Park:2019} proposed a Primary-backup optimization to reduce the exposed write latency for external clients, but 
its correctness relies on
commutative operations.
Howard's optimization~\cite{Howard:2019} allows Paxos to commit after 1 \CAP{RTT} in conflict- and failure-free rounds, albeit reads are not local.
In contrast, Hermes is not limited to commutative operations and affords local reads. 

\beginbsec{Reliable transaction commit}
Hermes provides single-key linearizabile reads, writes and \CAP{RMWs}, but does not offer fully reliable multi-key transactions. 
The distributed transaction commit requires an agreement on whether a transaction should atomically commit or abort: the transaction may only be committed if all parties agree on it. 
A popular protocol to achieve this is the two-phase commit (\CAP{2PC})~\cite{Gray:1978}. However, the \CAP{2PC} is a blocking protocol and must be extended to three phases (\CAP{3PC}) to 
tolerate coordinator failures~\cite{Guerraoui:1995, Guerraoui:2002, Skeen:1981}. 
A more common way to achieve reliable transactions is layering a transactional protocol over a reliable replication protocol~\cite{Zhang:2013, Kraska:2013, Corbett:2013}. For instance, FaRM and Sinfonia use Primary-backup~\cite{Dragojevic:2015, Aguilera:2007}.
In this latter setting, Hermes can be used as the underlying reliable replication protocol to increase locality and 
performance.

\beginbsec{Geo-replication}
Hermes is designed for replication within a local area network (e.g., a datacenter), where network partitions are rare. 
The conventional wisdom for fault-tolerant replication across dataceneters is to offer causal consistency which allows execution in all sites under partitions. 
A causal replication protocol could be tiered over several in\-de\-pend\-ent\-ly-replicated geo-distributed instances managed by Hermes (instead of \CAP{CR}~\cite{Lloyd:2011, Almeida:2013}) to accelerate geo-replication.

\vspace{-1pt}
\section{Discussion}
\vspace{-5pt}
\label{sec:discussion}
\beginbsec{Are local reads beneficial in a large-scale datastore?} \\
Throughout the paper, we report the latency of operations with respect to a
node (replica) in a distributed datastore.
In a large-scale datastore, clients might be external and not co-located with a replica they desire to access. Although in this case, reads in Hermes do not provide locality with respect to the client, they still ensure load balance and low latency. This is because, 
in Hermes, a remote read from an external client would be solely served by just one replica without 
additional messages, delays or coordination amongst replicas.

\beginbsec{Reducing write latency of external clients}\\
For the protocols discussed in this work, if clients are external, an additional round-trip is required to reach and get a response from the replica ensemble. Thus, the common-case exposed latency for an external client to commit a write in Hermes is 2~\CAP{RTT}s. To reduce the response time,
followers can send \CAP{ACK}s to both the coordinator of the write and the client\footnote{Coordinator must send an \CAP{ACK} to the client as well.} (\textbf{F\textsubscript{\CAP{ACK}}}). This reduces the latency to complete linearizable writes from external clients to 1.5~\CAP{RTT}s.
The message cost of this optimization (about twice the number of \CAP{ACK}s 
of the baseline protocol) is linear with the replication degree.

\beginbsec{Hermes without Loosely Synchronized Clocks (LSCs)}\\
This paper considers
a failure model with \CAP{LSC}s. 
Hermes leverages \CAP{LSC}s only for the \CAP{RM} lease management, to ensure that a node with a lease always has the latest membership.
However, Hermes can be efficiently deployed in the absence of \CAP{LSC}s with minor modifications.
Hermes' writes seamlessly work without \CAP{LSC}s, since they commit only after all acknowledgments are gathered, which occurs only if the coordinator has the same membership as every other live follower\footnote{Followers with different membership value would have otherwise ignored the received \CAP{INV}s due to discrepancy in the message epoch\_ids (\S\ref{sec:reliable-protocols})}.

Linearizable reads in Hermes can also be served without \CAP{LSC}s. The basic idea is to use a committed write to {\em any} key after the arrival of a read request as a guarantee that the given node is still part of the replica group, hence validating the read. More specifically, observe that a node can establish that it is a member of the latest membership by successfully committing a write. Using this idea, a read at a given node can be speculatively executed but not immediately returned to the client. Once the node executes a subsequent write to any key and receives acknowledgments from a \textit{majority} of replicas, it can be sure that it was part of the latest membership when the read was executed. Once that's established, the read can be safely returned to the client. Note that a majority of acknowledgments suffices because the membership itself is updated via a majority-based protocol. 

If a subsequent write is not readily-available (\eg due to low load) the coordinator can send a \textit{membership-check} message which contains only the membership epoch\_id to the followers. The followers will acknowledge this message if they are in the same epoch. After a majority of acknowledgments is collected, the coordinator returns the read. The \textit{membership-check} is a small message and can be issued after a batch of read requests are speculatively executed by the coordinator.
Thus, although serving reads without \CAP{LSC}s increases the latency of reads until a majority of replicas respond, it incurs zero (if a subsequent write is timely) or minimal network cost 
to validate the read.

\vspace{-3pt}
\section{Conclusion}
\label{sec:conclusion}
\vspace{-2pt}

This work introduced Hermes, 
a membership-based reliable replication protocol that offers both high throughput and low latency. Hermes uses invalidations and logical timestamps to achieve linearizability, with local reads and high-performance updates at all replicas. In the common case of no failures, Hermes broadcast-based writes are non-conflicting and 
always commit after a single round-trip.
Hermes tolerates node and network failures through its safe write replays.
An evaluation of Hermes against state-of-the-art protocols shows that it achieves superior throughput at all write ratios and considerably reduces tail latency.

\vspace{-2pt}
\begin{acks}
We thank our shepherd, Rodrigo Rodrigues, and our anonymous reviewers for their constructive comments and feedback. 
This work is supported by Microsoft Research and ARM through their PhD Scholarship Programmes, as well as EPSRC grants EP/M027317/1 and EP/L01503X/1.
\end{acks}

\newcommand{\showDOI}[1]{\unskip}
\newcommand{\shownote}[1]{\unskip}

\bibliographystyle{ACM-Reference-Format}
\bibliography{references}

\end{document}